\documentclass[journal]{IEEEtran}
\ifCLASSINFOpdf
\else
\fi
\usepackage{color,graphicx}
\usepackage{epstopdf}
\usepackage{setspace}
\usepackage{multirow}
\usepackage{hyperref}
\usepackage{cite}
\usepackage{amsmath}
\usepackage{amssymb}
\usepackage{colortbl}
\graphicspath{{Figures/}}
\definecolor{Gray}{gray}{0.90}
\DeclareMathSizes{10}{9}{7}{6}
\IEEEaftertitletext{\vspace{-2\baselineskip}}

\begin{document}
\title{Towards a Green and Self-Powered Internet of Things Using Piezoelectric Energy Harvesting}
\author{{Mahyar~Shirvanimoghaddam, Kamyar~Shirvanimoghaddam,~Mohammad Mahdi Abolhasani, Majid Farhangi, Vahid Zahiri Barsari, Hangyue Liu, Mischa Dohler, and Minoo Naebe}

\thanks{M. Shirvanimoghaddam is with the Centre for IoT and Telecommunications, School of Electrical and Information Engineering, The University of Sydney, NSW, Australia (email: mahyar.shm@sydney.edu.au).

K. Shirvanimoghaddam and M. Naebe are with The Institute for Frontier Materials, Deakin University, VIC,  Australia (email: \{kshirvani, minoo.naebe\}@deakin.edu.au).

M. M. Abolhasani is with Chemical Engineering Department, University of Kashan, Kashan, Iran (email: abolhasani@kashanu.ac.ir).

M. Farhangi, V. Zahiri Barsari, and H. Liu are with the School of Electrical and Information Engineering, The University of Sydney, NSW, Australia (email:\{mfar8291, vzah5643, hliu7760\}@uni.sydney.edu.au).

M. Dohler is with the Centre for Telecommunications Research, Department of Informatics, King's College London, UK (email: mischa.dohler@kcl.ac.uk).

$\copyright$ 2019 IEEE. Personal use of this material is permitted. Permission from IEEE must be obtained for all other uses, in any current or future media, including reprinting/republishing this material for advertising or promotional purposes, creating new collective works, for resale or redistribution to servers or lists, or reuse of any copyrighted component of this work in other works.}}

\maketitle

\begin{abstract}
Internet of things (IoT) is a revolutionizing technology which aims to  create an ecosystem of connected objects and embedded devices and provide ubiquitous connectivity between trillions of not only smart devices but also simple sensors and actuators. Although recent advancements in miniaturization of devices with higher computational capabilities and ultra-low power communication technologies have enabled the vast deployment of sensors and actuators everywhere, such an evolution calls for fundamental changes in hardware design, software, network architecture, data analytic, data storage and power sources. A large portion of IoT devices cannot be powered by batteries only anymore, as they will be installed in hard to reach areas and regular battery replacement and maintenance are infeasible. A viable solution is to scavenge and harvest energy from environment and then provide enough energy to the devices to perform their operations. This will significantly increase the device life time and eliminate the need for the battery as an energy source.  This survey aims at providing a comprehensive study on energy harvesting techniques as alternative and promising solutions to power IoT devices. We present the main design challenges of IoT devices in terms of energy and power and provide design considerations for a successful implementations of self-powered IoT devices. We then specifically focus on piezoelectric energy harvesting and RF energy harvesting as most promising solutions to power IoT devices and present the main challenges and research directions. We also shed light on the security challenges of energy harvesting enabled IoT systems and green big data.
\end{abstract}

\begin{IEEEkeywords}
Energy harvesting; Internet of Things (IoT); RF energy harvesting; piezoelectric.
\end{IEEEkeywords}
\IEEEpeerreviewmaketitle

%---------------------------------------------------------------------------

% energy harvesting provides sustainable and independent operation with very long life-time but usually with unregulated power flow.
%The design of new generation systems should take into account of the limitations of energy harvesting, e.g., scarcity, unregulated flow and non-availability of power in some time intervals which can not be predicted beforehand.

\section{Introduction}
%\IEEEPARstart{E}{nergy} is a valuable resource in wireless communication, navigation and sensor nodes .The vast reduction in size and power consumption of CMOS circuitry has led to a large research effort based around the vision of ubiquitous networks of wireless communication nodes. As the networks, which are usually designed to run on batteries, increase in number and the devices decrease in size, the replacement of depleted batteries is not practical. Methods of scavenging ambient power for use by low power wireless electronic devices have been explored in an effort to make the wireless nodes and resulting wireless sensor networks indefinitely self-contained.

\PARstart{R}{ecent} advancements in miniaturization of devices with higher computational capabilities and ultra-low power communication technologies are driving forces for the ever growing deployment of embedded devices in our surroundings. This will transform every physical object, i.e.,  \textit{thing}, into an information source with the potential to communicate with every other thing in the network. This ecosystem of connected things are called \emph{Internet of Things (IoT)} which aims at transforming the world into a fully digital and inter-connected world.

%Gartner estimated that by 2020 over 50 billion devices will be attached to the communication networks and  [Gartner]. Most of these devices will be wireless due to the expense, inconvenience, and in some cases infeasibility of wiring them. They also have strict size, power, and service constraints, and lifetime.

IoT applications and services cover almost any sector where embedded devices can replace human in performing tasks.  Examples are home automation, agriculture, smart cities, industrial automation, healthcare, remote monitoring, and many more.  IoT provides a network of connected devices that real time information can be shared and used in order to enhance life quality, improve industry processes, energy efficiency, and level of services. IoT significantly improves supply chain efficiencies and develop new services for retailers \cite{IntelIoT}. Such a network of inter-connected devices  enables the factories to get humans and enterprise systems more involved with the machines in the whole supply chain system. This will improve the total revenue and improve customer satisfaction, and bring the customer experience into a whole new level. %Fig. \ref{fig:smartworld} shows a wide range of IoT applications and services.
%\Figure[t!](topskip=0pt, botskip=0pt, midskip=0pt)[width=0.98\columnwidth]{IoT}
%{IoT Sectors. The information of this figure has been obtained from \cite{ResearchIoT}.\vspace{-5ex}\label{fig:smartworld}}

Gartner estimated that the total number of IoT devices will reach 20 billion by 2020 \cite{Gartner}. Only a small fraction of these devices will be our smart phones and computers and a majority of them perform simple tasks such as sensing and transmitting a small amount of data once in a while. Major industries will be equipped with embedded smart sensors to gather data from their machinery to better track inventory and manage machines. This will significantly increase efficiency, revenue, and save costs and even lives. There are about 4.7 million developers in the world who can create IoT devices, and the number is increasing at a rate of around 3\% per year \cite{Gartner}. Intel reported that IoT technology will worth about USD 5.2 trillion globally, where healthcare and manufacturing will share USD 2.5 trillion and USD 2.3 trillion, respectively \cite{IntelIoTGuide}.

Devices in IoT are usually divided into two main categories. \textit{Basic devices} only provide basic services of sensor reading and/or actuation tasks,  and some limited support for user interaction. They have basic processing and communication capabilities with limited memory and are usually running with small batteries. Basic devices are usually used in \textit{massive IoT} applications,  such as consumer electronics and smart metering. In massive IoT, devices must have very low power consumption and the network should provide long coverage. Also due to the massive scale of the system, the devices should be of low cost and maintenance. On the other hand, \textit{advanced devices} are capable of running complex tasks and usually support cellular and wide area network connections. Examples of advanced devices are smart phones and laptops. These devices are widely used in \textit{critical IoT} applications, with high demands for reliability, availability, and low latency.

Powering the IoT devices is a major challenge, which becomes crucial with the rapid development of relative technologies \cite{R11}. In fact in many IoT applications, devices need to be powered in a self-sufficient and sustainable fashion. Most of the devices will be battery operated due to cost, convenience, size, and the fact that they are implemented in hard-to-reach areas. In many IoT applications, long life time of the devices is of prominent need, as the battery maintenance is not feasible due to cost, inconvenience, and the size of the networks. Recent studies showed that more than 3 billion batteries are discarded in the USA every year \cite{jayakumar2014powering}, and the penetration of IoT technologies in every sector will exacerbate this problem.

The trend so far to improve energy efficiency and propose viable solutions for long-life time devices for IoT has been mainly focused on three main directions. First, is to reduce the energy consumption of every component of the device in any operation mode, including embedded sensors, microprocessors, and transceivers. Second, is to increase battery efficiency and have smaller yet more efficient batteries. Third, is to increase the device life time by introducing duty cycling and event-driven communication to reduce the energy usage. While in many cases these trends have achieved significant improvements, there are still major challenges which have not been solved yet. In fact, the IoT applications which only rely on batteries require battery replacement; otherwise node failure will dramatically drop the system performance. Battery replacement is however not a feasible option due to cost and physical operations. 

Energy harvesting (EH) provides several solutions for powering IoT. EH techniques harvest energy from different energy sources in the environment, such as thermal, vibrations, solar, and radio frequency (RF) signal energy. This provides a potentially unlimited power supply for IoT devices and significantly increases the device life-time.   Energy harvesting market has experienced a huge growth from \$131.4 million in 2012 to  \$4.2 billion in 2019\footnote{\url{https://pitchengine.com/pitches/59354ec3-359e-4476-89ef-2ce639997a21}}. The main reasons behind this growth are 1) the demand for micro-power generation to charge thin film batteries, 2) recent advancements in energy storage devices, i.e., batteries and super-capacitors, 3) the efficiency of EH devices have significantly improved and 4) the fact that the price of  super-capacitors and thin-film batteries dramatically decrease. There are also significant cost reductions, which makes them suitable choices for a wide adoption in IoT applications. %The focus of the paper is on the enabling EH technologies for IoT applications.
Energy harvesting has been widely studied in the context of wireless sensor networks (WSNs) \cite{R15} and wireless body area networks (WBANs)  \cite{R16} in the literature. These studies have mainly focused on 1) energy harvesting technologies  \cite{chalasani2008survey,yildiz2009potential}, 2) power management mechanisms \cite{raghunathan2006design,pimentel2010power}, 3) challenges and practical issues of energy harvesting for WSNs \cite{seah2009wireless}, and 4) sensor node design \cite{sudevalayam2011energy}. However, due to unique requirements of emerging IoT applications and the differences between IoT and WSNs in general , there is a major need to study and investigate the EH techniques for the specific uses for IoT. In particular, IoT devices may interact with multiple network tiers to deliver their messages and receive feedbacks and act upon the command received from the network. Also, most studies on EH techniques for WSNs have focused on short-range communication technologies, which might be unsuitable for many IoT applications. In fact, several low-power wide area networks (LPWAN) have been proposed and implemented to provide long coverage and high capacity IoT services. The IoT Global Forecast \& Analysis 2015-2025, from Machina Research expects that 11\% of IoT connections in 2025 to use LPWAN connectivity technologies \cite{maschina}. This shows the need for the comprehensive study of self-powered solutions for emerging IoT applications which are enabled using LPWAN technologies. This paper reviews several energy harvesting techniques, provides insights on the design of suitable self-powered IoT devices, discusses pros and cons of each EH technique, provides some detailed discussions on the piezoelectric energy harvesting materials and design for self-powered IoT, and finally discusses the future challenges of self-powered IoT. The paper is of survey nature aiming at paving the path to a green and sustainable IoT ecosystem.

The remainder of the paper is organized as follows. In Section II, we briefly introduce Internet of Things and different applications and requirements. Section III provides a comprehensive overview of energy in IoT ecosystem, where we also discuss about IoT devices, energy harvesting use cases in IoT. In Section IV, different approaches to power IoT devices are studied and pros and cons for each approach are explained. In Section V, we focus on piezoelectric energy harvesting and provide a comprehensive study on piezoelectric effect, mode of use, and piezoelectric materials. In Section VI, we focus on RF energy harvesting for IoT and discuss about the challenges and research directions. Security challenges of EH-enabled IoT systems are discussed in Section VII. We also briefly explain the green big data concept and shed light on the challenges and future research directions. Finally, Section IX concludes the paper.

\section{An Overview on IoT}

%
%\begin{table*}
%\centering
%\scriptsize
%\caption{Wireless technologies for long-range low power IoT applications.}
%\begin{tabular}{|p{2cm}||p{2cm}|p{2cm}|p{1.5cm}|p{1.5cm}|p{1.5cm}|}
%\hline
%LPWAN Technology&Frequency Band&Data Rate&Range & Transmit power\\
%\hline
%\hline
%LoRa&169/433/ 868/915 MHz&300bps- 50kbps&5km  urban/15km rural&0.2-0.8W\\
%\hline
%SIGFOX&868/902 MHz&$<$300bps&10km urban/50km rural&0.5-0.8W\\
%\hline
%NB-IoT&LTE Band&32kbps UL 48kbps DL&35km&0.2-1W\\
%\hline
%LTE-M&LTE Band&1Mbps&35km&0.5-1W\\
%\hline
%EC-GSM&LTE Band&1Mbps&35km&0.5-1W\\
%\hline
%\end{tabular}
%\label{tab:longwireless}
%\end{table*}
%\normalsize %
Internet of Things has attracted enormous attention from both industry and research sectors. This is mainly because of huge opportunities that IoT will create in the near future. IoT goes beyond personal computers and mobile phones, and converts every physical object to an information source and connect them to the Internet or local area networks. IoT services are mainly categorized into \textit{massive IoT} and \textit{critical IoT}. In massive IoT,  countless devices will send data to the servers via Internet or local networks. The devices therefore should operate with low power, low complexity, and of course must be of low cost. Critical IoT has completely different requirements such as, high reliability, availability and low latency; therefore more complex devices are required. Remote health care and traffic safety and control are examples of critical IoT services.

%Table \ref{tab:wireless} and Table \ref{tab:longwireless} summarize available low power wireless technologies for IoT.

The third generation partnership project (3GPP) has identified 4 different application family types for IoT. These include \cite{3gppLPWAN}:
%\Figure[t](topskip=0pt, botskip=0pt, midskip=0pt)[width=1.6\columnwidth]{iotdevices.png}
%{Different types of IoT devices. \vspace{-6ex}\label{fig:iotdevice}}

\textbf{Type 1} is tracking, assisted living, remote health monitoring, wearables and bicycle tracking. They require 5 years battery life, which can be also supported by energy harvesting for battery recharging, medium coverage, and latency of about 30 seconds (lower latency of 5 seconds is required for some tracking applications). They also require high mobility support. Wearable devices, such as smart watches and fitness monitors, which have longevity requirement of several days as the number of them per user is small and they can be regularly recharged. Many IoT devices are set-and-fort devices, such as home security, with the longevity requirement of several years and a user may have dozens of them, so regular replacement of batteries is inconvenient \cite{jayakumar2014powering}. Some devices are battery-less and self powered, such as RFID tags and smart cards.

\textbf{Type 2} is industrial asset tracking, microgeneration, agricultural livestock and environmental near real time monitoring. They require 5 to 10 years battery life time, latency of less than 1 seconds, and medium coverage.

\textbf{Type 3} mainly requires deep indoor coverage for smart metering, smart parking, smart building, home automation, and industrial machinery control. It also requires extended outdoor/rural coverage, for smart cities, waste management, agricultural stationary asset monitoring, and environmental sensor and data collection. They require very long battery life of 10 to 15 years, extended coverage, latency of 10 to 60 seconds, and low mobility support. Many of these devices are semi-permanent, which are deployed in bridges, buildings, and infrastructures for monitoring purposes, and expected to work for more than a 10 years. Regular replacement of batteries is infeasible.

\textbf{Type 4} is usually main powered, such as smart city lighting, home appliances and vending machines which are stationary. Low mobility support, indoor and outdoor coverage, and latency of less than 30 seconds are required. For vending machines latency of about 1 second is required.

Type 1, 2 and 3 are battery powered and require the battery life-time of 5 to 15 years. In Type 1 applications, the battery can be easily recharged, but for other types battery replacement/maintenance is an issue. 3GPP has also identified five challenges for the vast deployment of IoT services: \emph{1) device cost}, \emph{2) battery life}, \emph{3) coverage}, \emph{4) scalability}, and \emph{5) diversity} \cite{MagMahyar}. Most of IoT devices need small-sized and high-energy density batteries for longer lifetime. This calls for major technological improvements in battery development. Energy harvesting (EH) techniques are interesting alternatives to batteries, which promise to enable autonomous and deployable IoT applications, in energy-rich environments \cite{prasad2014reincarnation}. Extracting energy from the environment, enables the devices to reincarnate once they have accumulated enough energy from the ambience. Energy harvesting is then become an excellent choice for applications, which require increased lifetime, battery-less functionality and ease of maintenance \cite{IoTIEEE}.

\section{Energy in the IoT Ecosystem}
\begin{figure}[t]
  \centering
  \includegraphics[width=0.8\columnwidth]{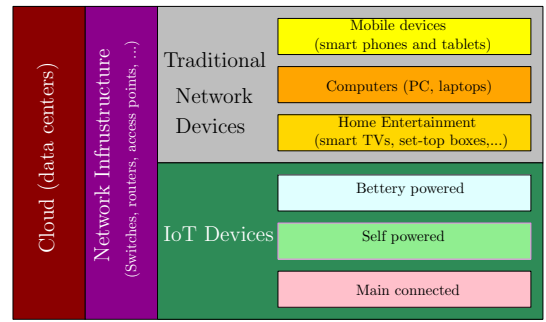}
  \caption{Energy consumers in the IoT ecosystem.}
  \label{fig:energyEco}
\end{figure}

IoT consists of several layers of devices, ranging from tiny devices to giant data centres. Fig. \ref{fig:energyEco} shows some of energy consumers in the IoT ecosystem. In the large scale, there are data centres and network infrastructure, which consume huge amount of energy; expecting to dramatically increase in near future. In the small scale, there are traditional network devices, such as mobile phones and tablets, computers and home entertainment, and IoT specific devices, which are somehow new to the market and are expected to grow exponentially in terms of the numbers. Only a small fraction of these devices are powered by connecting to the main power, and most of them would be battery operated or self-powered and use energy harvesting technologies. These were enabled through technological advancements in miniaturization of electronic devices, low power processing, low power wireless communications, and battery and energy storage, and the continuing fall in their prices and sizes. In what follows, we explain the unique characteristics of IoT systems, which distinguish them from WSNs, that have opened new challenges to be solved to enable sustainable IoT systems and applications.
%\Figure[t](topskip=0pt, botskip=0pt, midskip=0pt)[width=0.9\columnwidth]{EnergyEco.eps}
%{Energy consumers in the IoT ecosystem. \vspace{-6ex}\label{fig:energyEco}}

\subsection{Characteristics of IoT Energy Sources}
%\begin{figure*}
%  \centering
%  \includegraphics[width=1.9\columnwidth]{iotdevices.png}
%  \caption{Different types of IoT devices.}
%  \label{fig:iotdevice}
%\end{figure*}

%Here, we list these unique challenges of IoT systems and devices which must be carefully considered when designing the devices and communications protocols.
\subsubsection{Scalability} The energy source for IoT devices must be \textbf{scalable}. For the vast deployment of IoT services and applications, the devices need to be placed in all kinds of locations to collect data and communicate with the gateways. The devices may be located in hard-to-reach areas, so they need to work autonomously without human intervention. They also require minimum maintenance; therefore, batteries are not viable solutions.
\subsubsection{Maintenance-free}  The energy source for most IoT applications must be \textbf{maintenance-free}. IoT use cases usually involve a large number of devices. Connecting these devices through wiring to the main power is not feasible, as it restrict device movements and also increase the total deployment cost. Batteries are not feasible either as regular battery maintenance is impractical due to the large number of devices, the cost associated with batteries and also the enormous scale of maintenance expenses.
\subsubsection{Mobility support} Energy sources for many IoT application must \textbf{support mobility}. Many IoT devices are mobile, and the constant movement of the devices must be carefully considered when designing the devices and power management systems.
\subsubsection{Long life-time} Many IoT applications require energy sources that supports \textbf{long life-time} with minimal maintenance. In some IoT applications, for example in structural health monitoring, embedded wireless sensors must be deployed inside the buildings, bridges, etc., and they are supposed to work for several decades. In these applications, both the energy storage and wireless connectivity must be optimized in order to maximize the lifetime of the device.
\subsubsection{Flexibility} IoT energy sources must be \textbf{flexible in size and capacity} due to wide variety of IoT applications and services. For example in health monitoring applications, a tiny device will be implanted inside the human body to sense vital information and send it to a personal device. This requires long-lifetime tiny batteries or energy-harvesting enabled batteries. Many other applications, such as parking meters, may be connected to main power or can benefit from large photovoltaic cells due to their size and locations.
\subsubsection{Low-cost} Many IoT applications will require a large number of devices to be installed. These devices, and accordingly their power sources, should be of low cost; otherwise the application will provide low revenue or is very expensive, which limit its popularity.
\subsubsection{Sustainability} IoT will include trillions of devices, in different scales, and all of them consume power, which will affect our environment. IoT power solutions must be sustainable to avoid the depletion of natural resources in order to maintain an ecological balance. This further emphasizes the importance of energy harvesting power sources for IoT as alternatives to conventional non-environmentally friendly power sources.
\subsubsection{Environment-friendly} The expansion of IoT services will negatively impact the environment due to a large number of battery discarding, e.g.,  more than 125,000 tons of batteries are discarded in USA every year. Discarded AA batteries would circle the earth six times which worsen the problem. Addressing this problem is of urgent priority and energy harvesting techniques provide several solutions.

\subsection{Powering IoT Devices}

\subsubsection{Main Power}
Devices in IoT, may be connected to a wired power supply. This is more suitable for IoT applications with fixed-location devices, where a constant power supply can be connected to the device through wires or cables. This however makes the devices immobile, which limits its application in massive IoT. Moreover, it is impractical to connect every device to the power supply through wires when the number of devices is very large. This option is only feasible when the number of devices is very small, and due to specific requirements of the devices is the only way to power the devices.% Examples include video surveillance and monitoring applications, where a specific area is constantly monitored by a few high-quality cameras and the data is continuously sent to the security units. Another major application is smart appliances, which are placed in fixed-locations in homes, so they can be connected to the main power.

\subsubsection{Battery and Super-capacitors}
Battery is the most common energy source which has been widely used in our everyday devices. The stored energy in batteries however is limited, therefore the battery-driven systems have a finite lifetime. It is also difficult and costly to regularly maintain and replace batteries especially when the nodes are remotely located or the number of nodes is very large. These are the main problems of battery-driven systems, which limit the use of batteries in some massive IoT applications.
%Lithium batteries are the most efficient batteries which have the highest power densities and efficiencies which can provide higher battery lifetime, suitable for some IoT applications. However, massive IoT applications require the devices to be tiny and autonomous, which put strict limitations on the energy storage and power management of IoT devices. These make the batteries not viable solutions for them \cite{IoTIEEE}.
 Non-rechargeable batteries cannot be solely used for many IoT applications due to ecological implications and the fact that they have only limited storage \cite{IoTIEEE,El14IoT}.  %Imprint Energy \cite{ImprintIoT} developed 3D printed Zinc rechargeable batteries were developed by for powering IoT devices, which do not require heavy installation and can be formed into any shape; allowing for customized applications. These batteries are slim and flexible and customizations ensures the required capacity and voltage to avoid extra power conditioning \cite{IoTIEEE}. Another solution which has low power density and high energy density is solid-state thin-film batteries suitable for long-term deployment of IoT devices. The flexibility of these batteries and the fact that they can be manufactured in IC packages have made them suitable candidates for many IoT applications that target low cost and ting device implementation \cite{Cymbet}. Super-capacitors have been also considered to replace rechargeable batteries, which have unlimited charge-discharge cycle, but suffer from high self-discharge (up to 20\% per day) \cite{IoTIEEE}.

Battery storage is a mature technology when compared with energy harvesting technologies, and the fact that batteries are available in different sizes and shapes, make them strong candidate for many massive IoT applications, which are expected to operate with ultra-low power and have limited life time, up to 10 years. Therefore, the battery technology still plays an important role in the IoT ecosystem for many years. The unique requirements of many IoT applications, open new challenges for battery providers.

\subsubsection{RF Energy}
In RF energy harvesting, the electricity is generated as a result of magnetic inductive coupling effect \cite{safak2014wireless}. It is basically based on the induction of an open circuit voltage around the receive loop from a loop which carries a time varying current. The flux and the open-circuit voltage are mainly determined by the distance between the turns of the loops, the amplitude of the transmit loop current, and the dimension and distance between the loops  \cite{safak2014wireless}. The induced voltage at the recieve loop can be used to power a passive RFID tag or stored in a rechargeable battery. The voltage induced in the receive loop is approximately 0.5 V \cite{sudevalayam2011energy}.

%Currently, we use this technology in electronic ID tags and smart cards, which are embedded with passive electronic devices and will be triggered when they are exposed to nearby energy rich sources which are transmitting RF signals. Considering the vast deployment of the devices in massive IoT, this solution may not be scalable as the environment needs to be flooded with RF radiation to power the nodes. Such radiations of RF signals would probably presents health risks for human beings.

Wireless energy harvesting (WEH) has been considered for powering IoT devices in \cite{kamalinejad2015wireless} and improvements in terms of being wireless, availability of the RF energy, low cost and relatively easy implementation were shown. Sensor nodes which are powered by WEH usually consist of a transceiver and antenna element, a WEH unit which is responsible for scavenging RF energy and delivering a stable output power, a power management unit, and possibly an onboard battery. Recently, Freevolt proposed an innovative technology, called Low Energy Internet of Things (LE-IoT) devices, which can harvest RF energy from both short-range and cellular wireless networks, such 4G, WiFi and Digital TV \cite{FreeVolt}.

\subsubsection{Thermal energy}
A thermal energy generator (TEG) converts temperature differences into electrical energy. A TEG usually suffers from low efficiency (5-10\%) which limits its widespread adoption \cite{tervo2009state,zhao2010enhancing}. However it has a long life cycle and stationary parts. To extract the energy from a thermal source, a thermal difference is required; e.g, 30 degree difference in the temperature of hot and cold surfaces of the device in the room temperature, results in only up to 10\% conversion efficiency \cite{kausar2014energizing}.

\subsubsection{Solar and Photovoltaic Energy}
Photovoltaic (PV) is considered as one of the most effective EH techniques to power IoT devices due to its power density, efficiency, and the flexibility in terms of different output voltage and current \cite{wang2016storage,ULP}.  When sunlight is directed to certain semiconductor materials, solar energy will be converted into DC power. This is called the photovoltaic (PV) effect. A solar cell is usually composed of silicon and when it is stroke by sunlight with enough energy, the electron and holes are separated and using an input an output regulator, electrons start to move towards the load  \cite{Shaikh20161041}.  To control the charging current to a battery or super-capacitor a maximum power point tracking (MPPT) unit is necessary, which also maximize the efficiency of the PV cells.

\subsubsection{Mechanical Energy}
Electrical energy can be harvested from vibrations, pressure and stress-strain. Electromagnetic, electrostatic, and piezoelectric are three main mechanisms to generate electricity from mechanical sources  \cite{Shaikh20161041}. In electromagnetic  energy harvesting, the electric current is generated when a magnet moves across a coil. In piezoelectric materials, an electric potential is induced at the terminals of a piezoelectric material due to the polarization of ions in the crystal as a result of the strain. In electrostatic converters, the plates of a charged capacitor are pulled using the vibration, which then results in electrical energy due to the change in the capacitance. Piezoelectric energy harvesters has the highest energy density, that is higher energy can be produced for a given surface area, which is very important in micro scales, where most IoT devices are supposed to operate. Electrostatic mechanism requires separates voltage source and electromagnetic usually generate low voltages.

\subsubsection{Human Body}
Human body is considered as a rich environment to scavenge energy to power wearable electronics \cite{starner1996human,shenck2001energy}. Wearable devices are very important in health monitoring applications, where sensor nodes are deployed on or implanted inside the human body, which form a network called wireless body are network (WBAN). As the battery replacement for wearable devices is inconvenient for people and sometimes impossible in cases when the devices are deployed inside the human body, the sensor nodes in WBAN must have very long lifetime. Therefore, energy harvesting from human body is favorable in these applications \cite{Shaikh20161041}.

\begin{table*}
  \centering
  \scriptsize
  \caption{Energy sources and their energy harvesting potential.}
  \begin{tabular}{p{1.5cm}p{1cm}p{2cm}p{4cm}p{7cm}}
    \hline
  \rowcolor{Gray}Energy source &References & Power density &Characteristics&IoT Applications\\
  \hline
      \hline
Solar/PV&\cite{wang2016storage,ULP,Shaikh20161041,mathuna2008energy,li2015retro}&100mW/cm$^3$ \cite{kausar2014energizing}& Require exposure to light, low efficiency for indoor devices& \textbf{Smart Home/Office} sensing, e.g., smoke detector, gas sensor, temperature/humidity sensor, fire detection, 

\textbf{Smart Factory}, e.g., asset tracking using RFID tags, \textbf{Smart Health}, e.g., activity tracker, weight monitoring, smart clothes, dental health, emergency/fall detection, 

\textbf{Smart Infrastructure}, e.g, smart road light, V2X communication, 

\textbf{Smart Logistic/retail}, e.g., quality control, tracking,

 \textbf{Smart Agriculture}, e.g., animal/pest/irrigation monitoring, smart gardening, 
 
 \textbf{Smart Environment Monitoring}, e.g., water quality/flood monitoring, hazard detection  \\
\rowcolor{Gray}RF energy&
\cite{hawkes2013microwave,qi2016acoustic,kellogg2014wi,kamalinejad2015wireless,FreeVolt,safak2014wireless,van2018ambient,muncuk2018multiband,drozdenko2018hardware,cid2016scalability,mishra2015smart,mishra2015charging
}&40 mW/cm$^2$ at 10m \cite{kausar2014energizing}&Low efficiency for indoor and out of line of sight& \textbf{Smart Factory}, e.g., asset tracking using RFID tags,

 \textbf{Smart Health}, e.g., wearables,nutrition monitoring, 

 \textbf{Smart Infrastructure}, e.g., smart roads, 
 
 \textbf{Smart Logistic/Retail}, e.g., product tracking, quality monitoring, waste management  \\
Body heat&\cite{Shaikh20161041,R16,starner1996human,shenck2001energy}&60 mW/cm$^2$ at 5$^{o}$C \cite{kausar2014energizing}&Available only for high temperature difference, Easy to build using thermocouple&\textbf{Smart Health}, e.g., wearables, activity tracker, smart body scale, sloop sensor, nutrition monitoring, GPS tracker, long-term body monitoring, emergency/fall detection\\
\rowcolor{Gray}External heat&\cite{tervo2009state,chen2012nanostructured,satyala2014nano,ThermoLife,Shaikh20161041,lu2010thermal,leonov2007thermoelectric,wang2009realization,cuadras2010thermal,ferrari2006modeling,harb2011energy}&135 mW/cm$^2$ at 10$^{o}$C \cite{kausar2014energizing}& Available only for high temperature difference, Easy to build using thermocouple&\textbf{Smart Infrastructure}, e.g., smart road, 

\textbf{Smart Logistic}, e.g., tracking, quality tracking \\
Body motion&\cite{singh2012self,Shaikh20161041,R16,ramsay2001piezoelectric,zurbuchen2013energy,dagdeviren2014conformal,mitcheson2008energy,benasciutti2013energy}&800 mW/cm$^3$ \cite{kausar2014energizing}& Dependent on motion, High power density, not limited on interior and exterior&\textbf{Smart Health}, e.g., wearables, activity tracker, smart body scale, sloop sensor, nutrition monitoring, GPS tracker, long-term body monitoring, emergency/fall detection\\
\rowcolor{Gray} Vibration/Piezoelectric&  \cite{discenzo2006power,sterken2003electret,sazonov2009self,starner1996human,kymissis1998parasitic,park2010piezoelectric,kwon2012high,hwang2014self,roundy2003energy,kausar2014energizing}&4 mW/cm$^3$, 50 mJ/N \cite{kausar2014energizing}& Has to exist at surrounding, High power density, not limited on interior and exterior& \textbf{Smart Home/Office}, e.g., smart light switches, intrusion detection, 

\textbf{Smart Health}, e.g., activity tracker, smart body scale, sleep monitoring, smart clothes, pacemaker, electrical toothbrush, emergency/fall detection, 

\textbf{Smart Infrastructure}, e.g., road pricing, smart roads, V2X communications, 

\textbf{Smart Logistic/Retail}, e.g., asset tracking, quality tracking, fleet management, 

\textbf{Smart Agriculture}, e.g., animal/pest/irrigation monitoring, smart gardening, 

\textbf{Smart Environment Monitoring}, e.g., water quality/flood monitoring, fie detection, hazard detection\\
\hline
  \end{tabular}
  \label{tab:3}
\end{table*}
\normalsize
%\subsection{A Brief Comparison between different Energy Harvesting Techniques for IoT}

In Table \ref{tab:3}, we have briefly compared different energy harvesting techniques. Energy harvesting from environment is usually uncontrolled and unpredictable and in most cases the conversion efficiency is low. PV cells provides high power density, but they require constant exposure to light which limits their application in many IoT use cases. Temperature based energy harvesting is very limited and can be useful when high temperature difference is guaranteed. RF energy harvesting is also limited due to low efficiency in indoor environment and also in case of multipath. On the other side, mechanical energy harvesting, especially piezoelectric EH can achieve high power densities which is suitable for IoT applications where mechanical vibrations is constantly available.

%There are several problems with solar and PV energy harvesting technologies, which make them not suitable for the wide adoption in many IoT applications. First of all, the solar panel must receive enough light to be able to generate enough power for the device. Therefore, when there is no light or the light intensity is low and varying, the PV energy cannot be properly harvested and used. Second, to harvest solar or PV energy, a relatively large (compared to the device size) PV cell must be installed to harvest enough energy, even if the light intensity is large. This is maybe the biggest issue with the PV energy harvesting which limits its application in massive IoT, which mainly require tiny devices to be installed in different places, regardless of the light intensity. Third, when the light intensity is varying, a battery must be also considered to store energy and provide that energy to the device, when the light intensity is low. This is quite common in applications which are powered by solar cells, where the energy is stored in batteries during the day and then used at night for different applications.

% Energy management then plays an important role in controlling the storage and usage of the converted energy, which must also take into account the optimal design of the demand side. That is the energy usage should be carefully coordinated and optimized between supply and demand sides, including signal processing and communication architectures, sleep scheduling, energy-efficient communication protocols and adaptive coding and modulation.

In the rest of the paper, we will focus on vibrational energy harvesting as the promising solutions for massive IoT applications. Piezoelectric materials show high power density and recent technological advancements in developing new materials with enhances electrical properties make them very attractive energy sources for the Internet of things applications in vibration-rich environments.
\begin{figure*}[t]
  \centering
  \includegraphics[width=1.7\columnwidth]{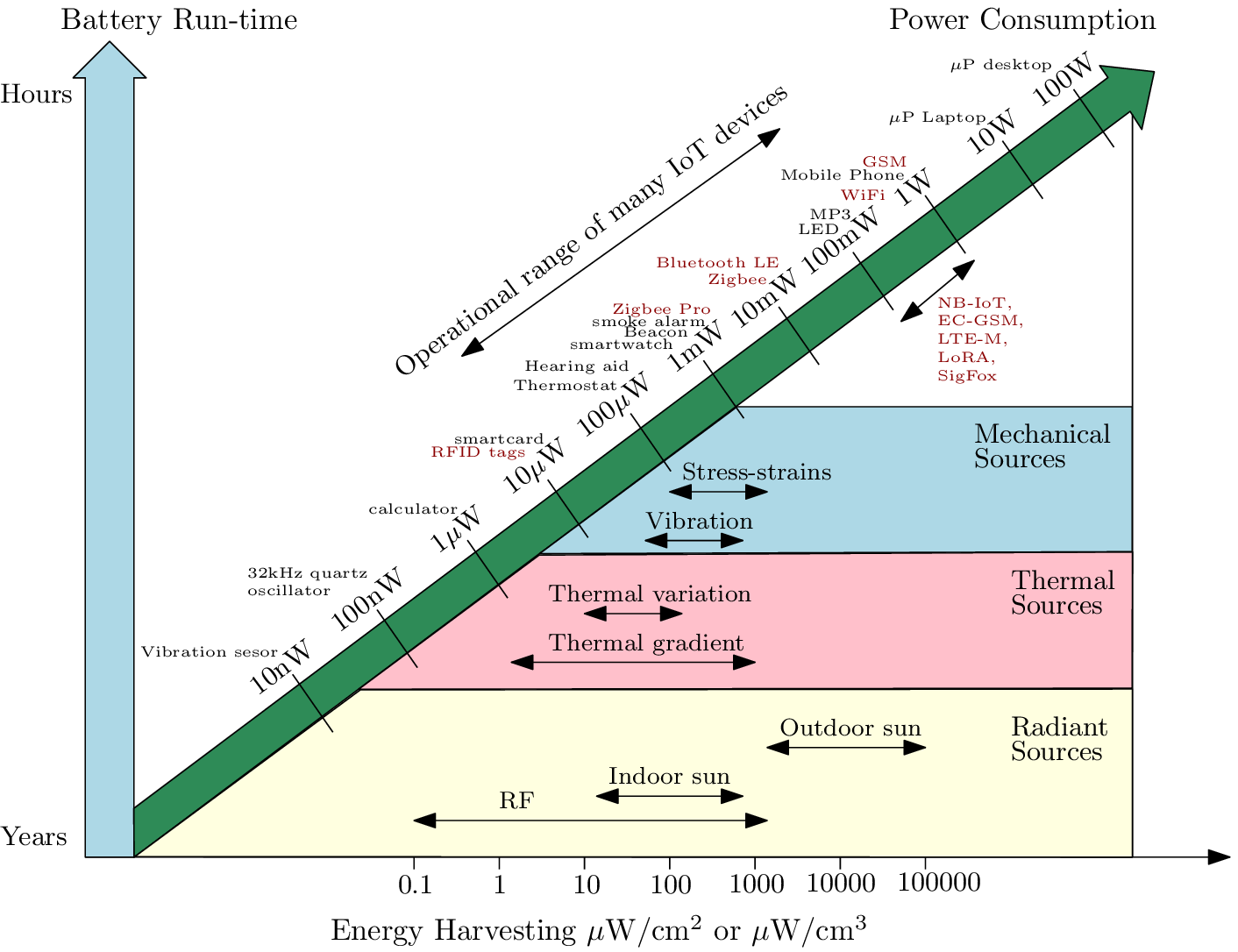}
 \caption{Power consumption for various applications and power densities for various energy sources (data obtained from  \cite{DSRC}). Power requirement for wireless technologies. }
 \label{fig:powerdens}
\end{figure*}
%\Figure[t](topskip=0pt, botskip=0pt, midskip=0pt)[width=1.4\columnwidth]{powerdensities.eps}
%{Power consumption for various applications and power densities for various energy sources (data obtained from  \cite{DSRC}).  \vspace{-6ex} \label{fig:powerdens}}
\subsection{Energy Harvesting Use Cases in IoT}
EH techniques use different sources of energy in the surrounding environment to harvest enough energy which is used later by the device for sensing, actuating, and communicating with the server. Solar, thermal, vibration, RF signals, and human body are only few examples of the available energy sources in the environment commonly used for energy harvesting, There are several scenarios where EH technologies can significantly enhance the system wide performance, in terms of energy, network life time, cost and maintenance. We divide these scenarios into three categories as detailed below.

First, energy harvesting is mostly useful in applications where devices are deployed in hard-to-reach areas, therefore, replacing the batteries for sensor nodes is almost impossible \cite{tanaka2014energy}. Wireless sensor networks (WSNs) have been widely studied for structural health monitoring, where the damage in aerospace \cite{becker2009autonomous}, buildings \cite{torfs2013low}, bridges \cite{whelan2009highway}, and mechanical infrastructures is detected by sensor nodes. The goal is to replace qualitative visual inspection and time-based maintenance procedures with a more autonomous condition-based damage assessment processes \cite{park2008energy}. The aircraft health monitoring system using WSNs and energy harvesting techniques, including thermoelectric and vibration sources, was studied in \cite{bai2004wireless}. Energy harvesting using piezoelectric and inductive devices was also proposed to monitor the health of a real railroad track in \cite{nelson2008power}. Another example where energy harvesting is crucial is body sensor networks, where the sensors are required to harvest energy for autonomous operation  \cite{alemdar2010wireless,mascarenas2009mobile,he2009mems,toh2014autonomous}.

Second, energy harvesting can be used in applications which usually require too many devices and replacing their batteries is almost impossible or cost ineffective. Examples include electronic shelf labeling  \cite{de2010design}, body sensor networks, and massive IoT applications.

Third, in many applications there is no steady supply of electricity available. An example was discussed in \cite{prabhakar2011novel}, where energy harvesting was used in an agricultural  setting to enable a delay tolerant wireless sensor network.

In many IoT applications, such as smart home or smart offices, intelligence transportation systems, smart grids, and industrial monitoring, a large number of devices will be installed everywhere, sometimes in hard-to-reach areas. These devices will be the major consumer of energy in near future due to the rapid growth of their numbers, and the use of energy harvesting will decrease our dependencies on fossil fuels and other traditional energy sources which are depleting very fast. Energy harvesting can also promote environment-friendly, clean technology that saves energy and reduces $\mathrm{CO_2}$ emissions, which is a promising solution for achieving the next generation smart city and sustainable society \cite{tanaka2014energy}.

On the other hand, energy harvesting could save lots of energy in a wider scope, as it could hugely reduce the cost of modifications in the buildings and industries for wiring and maintenances. Energy rich environment, such as industries and vehicles, must be capitalized on this available energy, where the small amounts of energy from the environment would be sufficient to run sensor nodes. Advanced sensing techniques in industries will significantly increase the performance and competitiveness, as constant monitoring through embedded devices reduces the cost and energy consumption associated with system failure and maintenance \cite{prieto2016self}. This however cannot be achieved if the system requires cables or if the battery have to be regularly replaced \cite{aygun2011wireless,hou2012novel}. Wireless solutions and energy harvesting techniques provide a wide range of solutions for the sensors to become maintenance free. Some examples are the use of piezoelectric energy harvesters for vibration monitoring \cite{hu2010vibration}, solar energy harvesting for autonomous field sensors \cite{decker2014solar}, and recently proposed multi-source energy harvester strategies for wireless sensor networks \cite{weddell2013survey}.

Fig. \ref{fig:powerdens} shows power requirements of different applications and the power densities for various energy sources. As can be seen selecting the energy source for IoT applications, depends mainly on the application and the specific requirements of that. Silicon Lab \cite{BatterySilicon} has recently identified 5 fundamental considerations for powering wireless IoT sensor products. These include, 1) the target market and its specific requirements on cost, reliability, and network lifetime, 2) energy efficiency and the choice of wireless connectivity, 3) the required transmission strength, duration and duty-cycle between active and sleep states, 4) the sensor node and its power requirement and cost, and 5) the space constraints and storage energy.

%\Figure[t](topskip=0pt, botskip=0pt, midskip=0pt)[width=1.4\columnwidth]{EnergySourcesIoT.eps}
%{Energy sources for IoT applications.  \vspace{-6ex} \label{fig:taxonomy}}
%\input{Section-2}
%\input{Section-3}
\section{Piezoelectric Energy Harvesting}
Piezoelectric energy harvesters provide the consistent source of energy and the potential of generating electricity from piezoelectric energy harvesters is higher than alternative energy harvesting technologies. In this section, we study piezoelectric phenomena to be able to compare existing piezoelectric materials in the market. Piezoelectric materials can be optimized based on the intended application, to deliver the required level of voltage or current, and can be manufactured in any shape or size.

%However, the challenge of converting the energy from broadband frequencies and harnessing a different number of sources of vibration at the wide variation in frequencies to produce a consistent supply of electric charge can hamper the growth in the piezoelectric energy harvesting market in the near future. Additionally, the key advantage offered by the piezoelectric material, that they can be optimized to as per the intended application and their ability to be manufactured in any shape or size, and wide ranging application of the technology is expected to augment the growth in the piezoelectric energy harvesting market.

\subsection{The Piezoelectric Phenomena}
\begin{figure}[t]
  \centering
  \includegraphics[width=1\columnwidth]{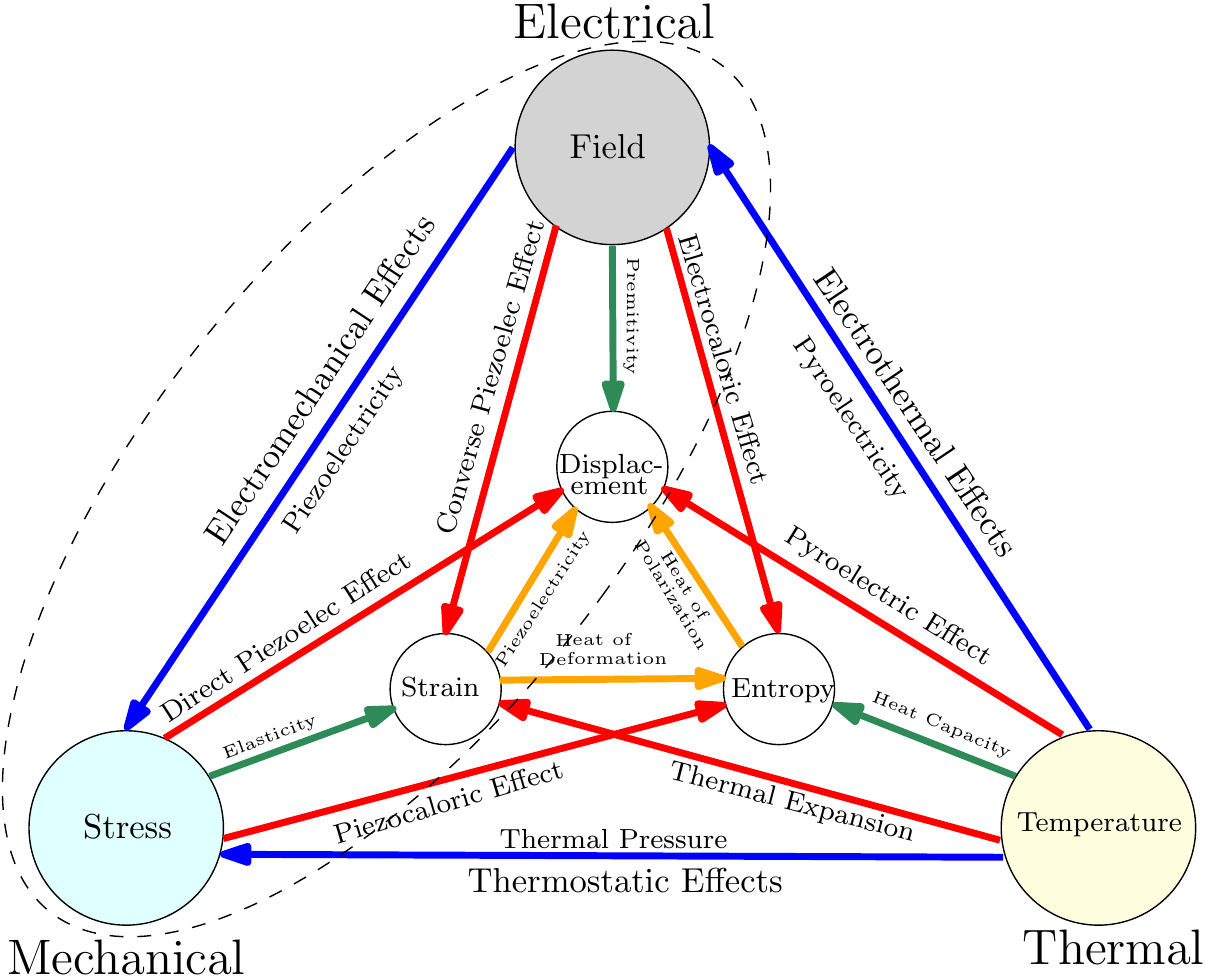}
  \caption{A Heckman diagram representing the interrelationship between mechanical, thermal  and electrical properties of materials \cite{K1}.}
  \label{fig:kam1}
\end{figure}

%The piezoelectric effect refers to electromechanical behaviour of some materials due to existence of stress or electrical field in them.
The Heckman diagram shown in Fig. \ref{fig:kam1} provides an overview of the interactions between mechanical, thermal and electrical properties of solids \cite{K1,K2}. A pair of coupled effects is joined by a line, indicating that a small change in one of the variables produces a corresponding change in the other. We are interested in the interaction between the elastic and electrical variables, which is separated by the dashed ellipsoid in Fig. \ref{fig:kam1}.

\begin{figure}[t]
  \centering
  \includegraphics[width=0.8\columnwidth]{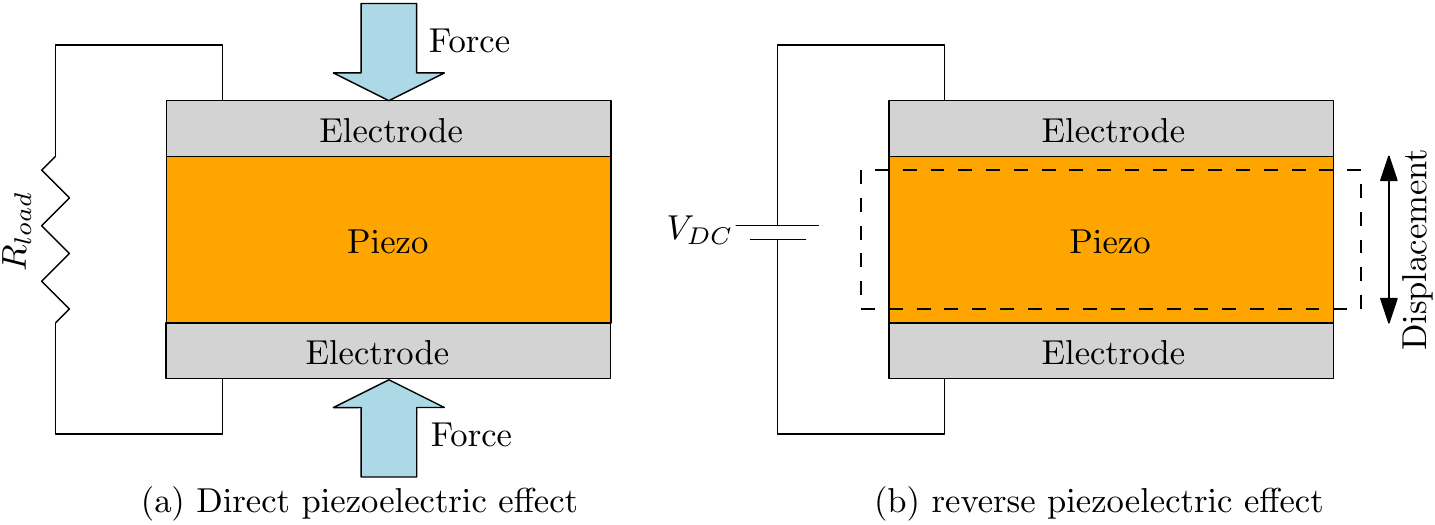}
  \caption{Direct and reverse piezoelectric effects.}
  \label{fig:piezoeffects}
\end{figure}

%\Figure[t](topskip=0pt, botskip=0pt, midskip=0pt)[width=0.97\columnwidth]{Heckman}
%{A Heckman diagram representing the interrelationship between mechanical, thermal  and electrical properties of materials \cite{K1}.\vspace{-6ex}  \label{fig:kam1}}
In 1880, Pierre and Jacques Curie measured surface charges that appeared on crystals of tourmaline, quartz, topaz, cane sugar and Rochelle salt when they  were subjected to an external mechanical stress, which is called the direct piezoelectric effect. In 1882, the inverse piezoelectric effect was confirmed by Jacques Curie, where the strain $S$ was observed in response to an applied electric field of strength $E$. Fig. \ref{fig:piezoeffects} shows direct and reverse piezoelectric effects. Piezoelectric materials show a small dimensional changes when exposed to an electric field, while some materials exhibit a reverse behaviour such that an electric polarization occurred when they are mechanically strained.

\begin{figure}[t]
  \centering
  \includegraphics[width=1\columnwidth]{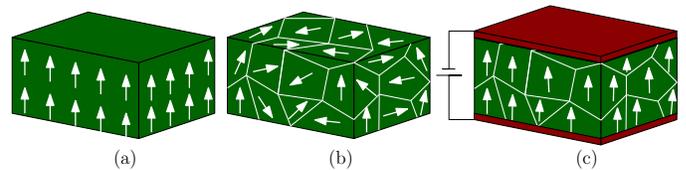}
  \caption{Different polarization in crystals. (a) Monocrystal, (b) Polycrystal, and (c) Polarization.}
  \label{fig:crystal}
\end{figure}

Direct piezoelectric effect is being considered as the main approach for harvesting energy from vibrations, where the external vibrations causes electrical charge on the terminal of the piezoelectric material \cite{brand2015micro}. The \emph{monocrystal} and \emph{polycrystalline} structure of same materials can be used to explain the piezoelectricity concept. As shown in Fig. \ref{fig:crystal}-a the polar axes of all carriers are aligned in the same direction in a monocrystal. In polycrystal (Fig. \ref{fig:crystal}-b) however, different regions within the material have different polar axes. The piezoelectric effect can be obtained by heating the polycrystal to the Curie point and then applying at a same time a strong electric field. The molecules can then move freely due to the heat which results in the re-arrangement of the dipoles due to the external field (Fig. \ref{fig:crystal}-c). When the material is compressed, a voltage will appear between electrodes.  The opposite polarity appears when the material is stretched (i.e., direct piezoelectric effect). On the other hand, when a voltage difference is applied the material will be deformed, and the material will vibrate when an AC signal is applied \cite{calio2014piezoelectric}.

%\subsection{Mode of Use an dFigure of Merit}
%As shown in Fig. \ref{fig:piezomode}, piezoelectric materials are commonly used in two different modes, i.e. $zz$ and $zx$. The first letter represent the voltage direction and the second letter represents the direction of the stress. Biomorphs are common type of $zx$ elements and made of two separate layers in such a way the when the top layer is in tension, the bottom layer is in compression or vice versa. If each layer is poled in the same direction (i.e., parallel polling) and electrodes are wired properly, the current produced by each layer will add. The layers can be also polled in opposite direction (i.e., series polling), therefore the voltages add. This concept can be used to manufacture bending elements with multiple layers, to increase the produced voltage or current. In theory, the poling and number of layers only affects the voltage to current ratio \cite{roundy2003energy}.

%Operation in $31$ mode leads to the use of thin bending elements in which a large strain in the $1$ direction is developed due to bending.
%\Figure[t](topskip=0pt, botskip=0pt, midskip=0pt)[width=0.8\columnwidth]{piezoeffects.eps}
%{Direct and reverse piezoelectric effects.  \vspace{-4ex}\label{fig:piezoeffects}}
%\Figure[t](topskip=0pt, botskip=0pt, midskip=0pt)[width=0.97\columnwidth]{CrystalPiezo.eps}
%{Different polarization in crystals. (a) Monocrystal, (b) Polycrystal, and (c) Polarization. \vspace{-6ex}\label{fig:crystal}}
%\begin{figure}[t]
%  \centering
%  \includegraphics[width=0.6\columnwidth]{modeofoperation.eps}
%  \caption{Illustration of $zz$ mode and $zx$ mode operation of piezoelectric material.}
%  \label{fig:piezomode}
%\end{figure}

%The coefficients for the direct piezoelectric effect can be described as follows. The  coefficient for the inverse effect is identical and  equal to  that of  the direct effect \cite{muensit2011energy}.
%\begin{align*}
%\text{Piezoelectric charge coefficient}d&=\frac{\text{dielectric displacement developed}}{\text{applied mechanical stress}}\\
%\text{Piezoelectric voltage coefficient} g&=\frac{\text{field developed}}{\text{applied mechanical stress}}\\
%\text{Piezoelectric stress coefficient} e&=\frac{\text{dielectric displacement developed}}{\text{applied strain}}\\
%\text{Piezoelectric stress coefficient} h&=\frac{\text{filed developed}}{\text{applied strain}}
%\end{align*}
%
%Let $\epsilon$ and $\beta$ respectively denote electrical permittivity and impermittivity, and $s$ and $c$ denote the elastic compliance and stiffness, respectively. The piezoelectric coefficients are then related by the following equations \cite{muensit2011energy}:
%\begin{align*}
%d&=\epsilon g = es\\
%e&=\epsilon h = dc \\
%g&=\beta d = h s \\
%h&=\beta e = g c
%\end{align*}
%
%The piezoelectric constitutive equations can then be written as follows \cite{muensit2011energy}.
%\begin{align*}
%  D&=eS+\epsilon^{S}E, \\
%  E&=-gT+\beta^{T }D,\\
%  E&=-hS+\beta^{S}D,\\
%  D&=dT+\epsilon^{T}E,
%\end{align*}

There are several figure of merit (FoM) for piezoelectric energy harvesting, but we only discuss about the power density which is more relevant to our topic. The power density, which is defined as the ratio of generated power over the active material volume or over the active material area. As shown in \cite{roundy2003energy}, the output power of a piezoelectric generator is proportional to the proof mass,  the square of the acceleration magnitude of the driving vibrations, and inversely related to the frequency. As mentioned in \cite{roundy2003energy}, piezoelectric devices provide high voltages and low currents. However, using multiple layers of biomorphs, it is easy to design a system that produces voltages and currents in an appropriate range.   %Electromechanical coupling, $k$, which depends on material shape, polarization axis, cut, static, or dynamic excitation, can be defined for both $zz$ and $zx$ modes of use. These couplings also depend on boundary conditions, which change with material shape and electrode position \cite{roundy2003energy}. To maximize the harvested power, the electrical damping needs to be increased, while the structural and aerodynamic damping need to be minimized \cite{roundy2003energy}.

\subsection{Advancements in Piezoelectric Fabrication for Energy Harvesting}
Various materials have been previously shown to demonstrate piezoelectric effect and used in several applications such as actuators, sensors, nanogenerators, atomic force microscopes (AFM), high voltage application, energy-harvesting devices, and medical applications. All of these technologies mainly rely on the mechanical energy harvesting  \cite{K2}. A wide range of both natural and synthetic materials exhibit piezoelectricity. More than 200 piezoelectric materials are available now, where for each energy harvesting application a specific type can be used. Many biological tissues such as bone, intestine and tendon exhibit piezoelectricity. Some naturally occurring crystals, such as quartz, rochelle salt, cane sugar, topaz, sucrose are classified as natural piezoelectric materials. Lead zirconate titanate (PZT), zinc oxide (ZnO), barium tintanate (BaTiO3), gallium orthophosphate(GaPO4), potassium niobate (KNbO3), lead titanate (PbTiO3), Lithiul tantalate (LiTaO3), langasite (La3Ga5SiO14), sodium tungstate (Na2WO3) and PVDF are the main synthetic piezoelectric materials, which have been widely used. %Table \ref{tab:2} shows some of the commonly used piezoelectric materials and their key properties.
%\begin{table}
%  \centering
%  \scriptsize
%    \caption{Comparison of promising piezoelectric materials.}
%  \begin{tabular}{|c||c|c|c|c|}
%  \hline
%Property&Units&PZT&PVDF&PZN-PT\\
%\hline
%\hline
%Strain coefficient ($d_{zx}$)&$10^{-12}$ m/v&320&20&950\\
%\hline
%Strain coefficient ($d_{zz}$)&$10^{-12}$ m/v&650&30&2000\\
%\hline
%Coupling coefficient ($k_{zx}$)&CV/Nm&0.44&0.11&0.5\\
%\hline
%Coupling coefficient ($k_{zz}$)&CV/Nm&0.75&0.16&0.91\\
%\hline
%Dielectric constant&$\epsilon/\epsilon_0$&3800&12&4500\\
%\hline
%Elastic modulus&$10^{10} N/m^2$&5.0&0.3&0.83\\
%\hline
%Tensile strength&$10^{7} N/m^2$&2.0&5.2&8.3\\
%\hline
%  \end{tabular}
%  \label{tab:2}
%\end{table}
%\normalsize

Compared to crystals, PVDF piezo films offer wider frequency range, sensitivity and mechanical toughness. PZT is suitable for low-noise application, while PVDF is most suited to the high frequency and large bandwidth applications. In addition to this, due to structural and physical property and lower density of the PVDF (1800gr/cm$^3$ compared to 7600gr/cm$^3$ for PZT) compared to  PZT, PVDF is a preferred candidate for lightweight sensors with curved surfaces and tiny complex structure \cite{K3,K4}.  Table \ref{fig:piezotable} shows the comparison of piezoelectricity performance of ceramic and polymeric materials at different conditions.
\begin{table*}[t]
  \centering
    \caption{Performance of piezoelectric materials at different geometries and loading conditions  \cite{K5}.}
  \label{fig:piezotable}
\centering
\scriptsize
\begin{tabular}{c|cccc|cc}
&\multicolumn{4}{c}{Ceramics}&\multicolumn{2}{|c}{Polymers}\\
\hline
\hline
Material&ZnO&ZnO&PZT&PMN-PT&PVDF&P(VDF-TrFE)\\
Device Length ($\mu$m)&200&10&5&150&100-600&60\\
Diameter (nm)&4000&100&500&200-800&600-6500&200\\
Device area (mm$^2$)&N/A&225&6&50&N/A&157\\
$V_{oc}$ (V)&0.035&35&0.7&6&0.01&3\\
\end{tabular}
\end{table*}
\normalsize
%$\eta_S$ (GJ m$^{-3}$)&2&1.5&7.8&1.8&0.18&0.27\\
%$\eta_T$ (PJ m$^{-3}$ Pa$^{-2}$)&1.3&1.7&4.1&13.8&89&44\\

%A variety of parameters has an effect on selecting ceramic or polymeric materials as a piezoelectric devices. This includes flexural modes, thermal shock resistance, fatigue and corrosion resistance, limits of mechanical deformation, and operating temperature range. As seen in Table \ref{fig:piezotable} and by  comparing the $\eta_T$ and $\eta_S$, it can be concluded that ceramics and polymers are more suitable for stress-driven and strain driven application, respectively.

A wide range of nanogenerators with different structural and functional properties has been developed to provide the optimum mechanical-to-electrical energy conversion by focusing on two main properties: piezoelectricity and triboelectricity \cite{K7}. Several inorganic materials in forms of nanorods and nanowires have been used for energy harvesting purposes (e.g nano PZT on PDMS polymer \cite{K8}, ZnO \cite{K6,K9,K10,K11}], GaN \cite{K12}, CdS \cite{K13}, and ZnS \cite{K14}). The main challenge in the use of inorganic ceramic piezoelectric are the brittleness and low formability compared to other materials. On the other hand, polymeric based nanogenerators exhibit outstanding processability, ease of fabrication and excellent formability, which make them ideal choices for energy harvesting applications \cite{K15,K16,K17,K18}. The PVDF and its copolymers are the most attractive semicrystalline polymers being used for piezolectrical purposes. %Five crystalline polymorph of PVDF are consisting of alpha (TGTG’), beta (TTTT), and gamma (TTTGTTTG’) phases, depending on the chain conformations as trans (T) or gauche (G) linkages. PVD based films demonstrate ferroelectric, piezoelectric and pyroelectric properties after poling \cite{abolhasani2014new,abolhasani2015different}. %Various  crystalline polymorphs of PVDF are shown in Fig. \ref{fig:kam2}.

\begin{figure}
  \centering
  \includegraphics[width=0.9\columnwidth]{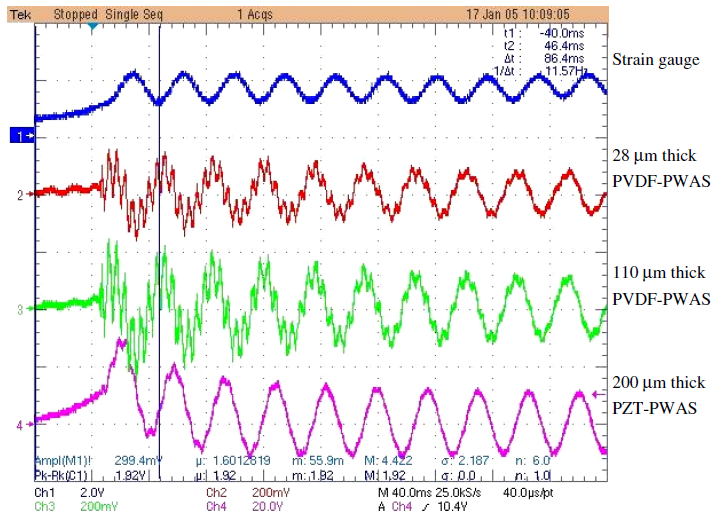}
  \caption{Vibration signal recorded by strain gauge, PVDF-PWAS and PZT-PWAS \cite{lin2006modeling}.}
  \label{fig:X1}
\end{figure}

In a research that carried out to compare the piezoelectricity of the PZT based and PVDF based sensors, the generated electric signals were recorded to analyse the frequency contents of the signals. The signals spectrum  are shown in Fig. \ref{fig:X1}. As can be seen the first three natural frequencies are $f_1 = 29.7$ Hz, $f_2 = 181$ Hz, and $f_3 = 501$ Hz. As can be seen although PZT based sensor deliver higher voltages compared to PVDF based sensors, they are less responsive to the higher frequencies \cite{lin2006modeling}.

\subsection{Design considerations for Piezoelectric-based EH-enabled IoT devices}

%Piezoelectric nanogenators have not been widely used for massive IoT. The main reason is that the can provide electricity when there is enough mechanical vibration in the environment. The generated power is also unpredictable. However, they can be used as a supplementary technology to recharge the battery. The battery will remain the main supplier of energy to the device and the piezoelectric nanogenerator help recharging the battery. This is feasible for many IoT applications that have low data rate demand and perform very limited tasks. These devices can work with a single battery charge for 5-10 years and that is more than enough to harvest energy from the environment to keep charging the battery. The aim is to minimize the battery replacement and maintenance and piezoelectric nanogenerators are excellent options for this purpose. 
%
%These is no comprehensive study on the lifetime of IoT systems equipped with piezoelectric nanogenerators for battery recharging. Unlike the RF energy harvesting that the energy harvesting medium will interfere with the information transfer medium which reduces the effective data rate, piezoelectric energy harvesting does not interfere with the information transfer and therefore does not limit the data rate. Further research in this area is required to better understand the dynamic of such systems so to be able to estimate the network lifetime and optimize the charging cycle of the device. 

We have conducted some research in the area of PVDF based piezoelectric materials. The primary focus has been on the investigation of fabrication methods for composite fibers and powder including nanomaterials.  By changing the fabrication techniques, process parameters, starting materials, the piezoelectric, morphological, and mechanical performance of PVDF are completely varied. By tuning the process parameters, tailored PVDF fibres with the desired piezoelectric performance can be developed.  To develop more efficient nanogenerator for energy harvesting from piezoelectric materials, we need to optimize the output performance to achieve a higher energy efficiency. In addition to piezoelectric performance, properties such as formability, corrosion/wear/fatigue resistance need to be considered in fabrication of flexible nanogenerators.

In what follows, we focus on the design considerations for piezoelectric energy harvesting IoT devices and provide some useful guidance in this regard. Some detailed design considerations for piezoelectric EH systems have been presented in \cite{roundy2003energy}.
\subsubsection{Unpredictable energy performance} Energy harvesting from the environment is unpredictable; that is the amount of power which is produced in a particular time instance is random which makes it difficult to continuously deliver sufficient power to the device. Harvesting sources also have low energy potential and low conversion efficiency \cite{AvnetUrl}. Therefore, there should be a power management module that balances the power generation with the power consumption. That is the load resistance should be chosen appropriately \cite{roundy2003energy}. 
\subsubsection{Power budget} Power budget \cite{AvnetUrl} is the most important factor when discussing the available options for powering IoT devices. That is what is the required voltage and current for the IoT device? How these requirements are changing  in different modes of operation, such as wake up, active, sleep or shut-down? As the largest power in IoT devices used when transitioning from deep sleep to active mode, the duty cycle or the update rate is also a crucial factor for choosing the right technique for powering the device. Minimizing the number of wake-up cycles and transmitting longer data bursts in each cycle is very important to increase the power efficiency \cite{Cypress}.
\subsubsection{Physical Damage} In a standard piezoelectric transducer which converts vibration to electricity, the printed-circuit boards of the electronics is subject to same vibration and can lead to premature failure. To avoid this, an accurate understanding of the generating environment, the power generated, and the time required, and the device power consumption and consumption time is required.
\subsubsection{Energy storage} As the current and voltage levels generated from energy harvesting sources system are fairly low, the batteries and super-capacitors must be designed to charge effectively at low power levels. Leakage and internal drain must be minimized too  \cite{AvnetUrl}. As shown in \cite{roundy2003energy}, the size of the storage capacitor should be at least 100 times the capacitance of the piezoelectric device. There is a tradeoff between the required capacitance for the piezoelectric EH systems and the demand side \cite{roundy2003energy}.
\subsubsection{Power IC} The power IC needs to carefully selected to match to the power generating element. That is the voltage/current/output characteristics of the power generating element output will differ according to the element, and it is necessary to choose a power IC that will provide optimal results \cite{AvnetUrl}.
\subsubsection{Output power} In piezoelectric biomorphs, a dramatic fall in the output power is observed if there is a mismatch between the resonant frequency of the converter and that of the driving vibrations \cite{roundy2003energy}. This has to be carefully considered when designing the IoT energy harvesting unit, so that the designed biomorph resonates at the frequency of the target vibrations. A magnets can be used to limit the vibrations of the device. However this solution is not viable as it adds bulk and cost and limits the broadband ability of the piezoelectric device. Pre-biasing the piezoelectric material is another approach to achieve a power gain up to 20 times.
\subsubsection{Oscillation proof mass} The power output of piezoelectric materials is proportional to the oscillating proof mass. In order to increase the power output, the mass should be maximized by taking into account the space limitations and also the yield strain of the piezoelectric material \cite{roundy2003energy}.
\subsubsection{Operational bandwidth} The output power of a piezoelectric energy harvester is inversely proportional to the bandwidth. The bandwidth is an important parameter when the piezoelectric harvesters is subject to unpredictable or uncontrollable ambient vibrations. The frequency of ambient vibrations is naturally uncontrollable, therefore narrow-bandwidth energy harvesters are impractical in most real applications.
%\Figure[t](topskip=0pt, botskip=0pt, midskip=0pt)[width=1.9\columnwidth]{CR}
%{RF-based energy harvesting in wireless cellular networks. \label{fig:cr}}

\subsubsection{Choosing the wireless technology}
Machina Research \cite{maschina} estimated that capillary or short-range wireless technologies will service more than 70\% of all IoT devices.  Most of these devices will be used in indoor environment and they have limited service demand and security requirements. From the market perspective, consumer electronics and building security and automation are the biggest short-range applications. It was also estimated that by 2025 more than 11\% of IoT connections will use LPWAN \cite{maschina}. Extended coverage GSM (EC-GSM), narrow band IoT (NB-IoT), and LTE for MTC (LTE-M) are three main solutions which have been proposed by 3GPP for massive IoT. These low power technologies were mainly designed to increase the battery life-time and the coverage, suitable for many IoT applications. Low power in this context refers to the ability of the IoT device to work for many years on a single battery charge. Long battery life-time is mainly achieved by defining new control and data channels and new power-saving and duty-cylcing functionalities for IoT applications \cite{MagMahyar}.

\begin{figure*}
\includegraphics[width=2\columnwidth]{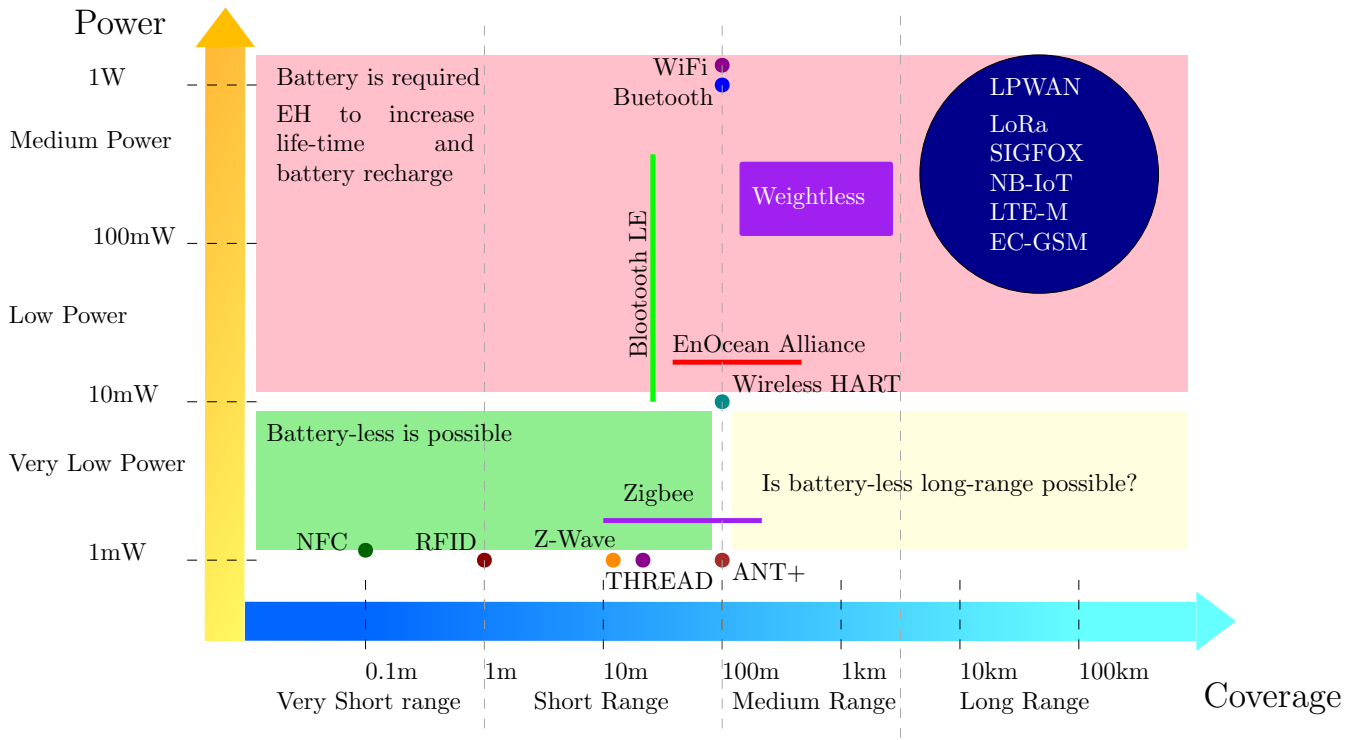}
\caption{Power consumption versus coverage for different wireless technologies. }
\label{fig:pc}
\end{figure*}

Fig. \ref{fig:pc} shows the power requirement for several wireless technologies and their coverage.  While short-range wireless communications operate over very low power which might be suitable for many IoT applications, they might not be feasible for applications that require long coverage. In fact, to have long coverage, the transmit power needs to be increased. Several low-power solutions have been proposed so far, which the transmit power ranges from 0.2 to 1 watt. This is much higher than what many short-range wireless technologies require, which may hamper the success of implementing self-powered LPWAN systems. In fact, due to the large transmit power of most LPWAN technologies, batteries will remain an essential parts of IoT devices operating over LPWANs. Energy harvesting techniques will then play a key role in increasing the device life-time by providing  a sustainable way to recharge the batteries.
%\Figure[t!](topskip=0pt, botskip=0pt, midskip=0pt)[width=1.8\columnwidth]{powercoverage}
%{Power consumption versus coverage for different wireless technologies. \vspace{-7ex} \label{fig:pc}}

\subsubsection{Delay Tolerant IoT Application}
In many IoT applications, the device transmit a small packet of data very infrequently. For example in water level monitoring applications, the water level is changing very slowly, therefore the data is updated only a few times a day. Therefore, the energy harvesting unit has enough time to harvest energy from the piezoelectric unit through the vibration.  In such systems, the periodic transmissions should happens such that there is enough energy available for the transmission. One need to adapt the sensing and transmission mechanisms to the environment conditions so that the energy stored in the given time interval is enough for the sensing and transmission. 

In many event-driven IoT applications, the device is usually remained in the sleep mode and wakes up only when an event happens. For example, an alarm based system that monitor the door movement. When the door is opened, the device needs to send an alarm signal to the gateway or central controller. The movement generate enough energy to be converted to the electricity by the piezoelectric unit. For such a system, the communication must happened in a timely manner as the device has a limited energy storage and cannot retransmit the message several times. Grant free access techniques is therefore necessary to minimize the access delay and power inefficiency due to several access requests by the IoT devices.

%Moreover, RF energy harvesting circuit and communication circuits are different, therefore, the received signal is usually splitted between the circuits or the supplied to only one circuit at a time. This however reduces the transmission rate. Using two separate circuits will add to the complexity and cost of the device as additional circuits such as a complex power management system may be required. 

%Several approaches have been proposed in the literature to mitigate this issues. 

%Power splitting at the receive may also reduce the effective transmission rate, as the communication circuit is needed to work under lower power as some power is used for energy harvesting. These issues have been considered in the literature and different simultaneous information and power techniques have been proposed to mitigate these issues.
\subsubsection{Delay Sensitive IoT Application}
In many IoT applications, the devices need to send the message in a timely  manner. Examples include tracking applications and health monitoring. For such applications, the device needs to be operated by batteries, so regular transmissions can occur in a timely manner. Piezoelectric modules are therefore necessary to recharge the battery to extend the device lifetime. 

As the device in transmitting in regular time intervals, the power management unit need to frequently switch between the two states, i.e., storage and usage. The design is therefore challenging in terms of maximizing the storage efficiency in the given time frame in order to minimize the probability of depleting the battery energy. 

A major issue is that the battery recharging cycle mainly depend on the current state of the battery. That is when the battery level is below 30\%, the battery could be recharged very quickly.

% \subsection{Users' Energy Fairness}

\subsubsection{Opportunistic Energy Harvesting using Piezoelectric Modules}
Piezoelectric materials are becoming cheaper nowadays and the fact that they provide a high energy density and can be produced in different shapes make them strong candidates for being used as back-up energy resources for many IoT devices. In particular, when the device is expected to work for a long time and exposed to many vibrations, the device could harvest enough energy using piezoelectric units and store it in the battery. This will significantly increase the device lifetime.

\subsubsection{User Mobility}
In many IoT systems, the devices may be mobile and therefore their channel to the base station is constantly varying. This means that the harvesting node may not harvest enough energy due to the movement and therefore cannot send its data. The problem becomes more challenging when the device carries critical information with a  strict latency requirement. 

%Most studies have focused on designing EH-enabled IoT systems where the devices have no mobility. In such scenarios, the system can be optimized to maximize the energy harvesting efficiency and network throughput. However, when the devices are mobile, an additional factor will play a key role in the overall network performance. This factor is the outage due to insufficient energy at the devices. While this might be temporary, it might change the whole system dynamic that can lead to inefficient radio resource and power usage. 

There is still a lack of comprehensive study on the dynamic of the IoT systems with mobile users powered by EH resources. RF-based energy harvesting seems a practical solution in mobile scenarios as the ambient RF energy is available everywhere. However, to maximize the systems performance the beamforming and radio resource allocation should be constantly updated due to changes in the channels. This will add extra communication  overhead and accordingly extra energy consumption. Further research in this area is required to better understand the outage performance of RF-based IoT systems and develop novel low-complexity algorithms for dynamic optimization of power and radio resource allocation.

\subsubsection{Hybrid Energy Harvesting for IoT}
As we discussed through this paper, energy harvesting techniques provide a sustainable way to power IoT devices. However, energy harvesting is not reliable and the device may not be able to store enough energy to power the circuits. One can use several energy harvesting techniques \cite{akan2018internet} to harvest energy from multiple sources at the same time. While this increases the energy storage efficiency, it potentially increases the complexity as multiple power management units should be installed to maximize the energy conversion for each energy source. Hybrid energy harvesting is an interesting technique especially now because of the advancements in energy harvesting techniques  miniaturization of sensors. 

\subsubsection{Security Challenges of EH-enabled IoT Systems}
There are three main limiting factors which might lead to several potential security risks. These are mainly caused by the environment and impact the EH-enabled IoT devices. Fist, energy harvesting from the environment is unpredictable; that is the amount of power which is produced in a particular time instance is random, which makes it difficult to continuously deliver sufficient power to the device. Harvesting sources also have low energy potential and low conversion efficiency \cite{AvnetUrl}. Although the power management module and the effective use of battery or super-capacitors can partly solve this problem, this may lead to temporary security risks due service unavailability. 

Second, the physical environment is very important in terms of reliability, possible hazards and incidents. In fact the device and EH circuit must be robust to real-world disruptions \cite{AvnetUrl}. 

Third, the physical property which is used to harvest energy from, can damage the associated electronics. For example, in a standard piezoelectric transducer which converts vibration to electricity, the printed-circuit boards of the electronics is subject to same vibration and can lead to premature failure. To avoid this, an accurate understanding of the generating environment, the power generated, and the time required, and the device power consumption and consumption time is required.

These limiting factors are potential sources of threats to the EH-enabled IoT systems. EH-enabled devices are vulnerable to malicious attacks that mainly limit accessing the energy resources. For example, IoT devices that are powered by PV cells may not receive enough sunlight, due to the blockage by a third-party, to perform their operations.  Also, the target device might not be easily compromised, but the attackers could easily change other devices behavior or the surrounding environment, which have interdependence relationship to achieve their aims \cite{zhou2018effect}.

% As a result, this feature could be maliciously used to reduce the difficulty of direct attack the target devices and bypass original defense mechanism. For example, back to the scenario described as the first example in the last paragraph, the hackers do not need to attack the automatic window control or thermometer.  However, he could compromise the smart plug that connected to the public network to turn off the air-conditioner in a room and trigger a temperature increase, which would result in the windows to open and create a physical security breach \cite{zhou2018effect}.

several papers in literature have studied the vulnerabilities, attacks and information leaks, and the design of security mechanism to provide security and privacy within the context of users and the devices. For a nice summary of these studies refer to \cite{li2016internet,lin2017survey,fu2017safety,roman2013features,sicari2015security,yang2017survey}. These studies are however in their initial stages and lack applicability, and many problems remain open \cite{zhou2018effect}. In particular, there is no unified framework to design a secure wireless systems based on EH techniques. When the devices are powered by EH sources, they are faced with an additional threat from the attacker who can change the environment. For example the RF source could be blocked and the devices cannot send their data anymore. In fact, the more the devices are geared with the environment, the more they are  vulnerable to the threats.

%This is an strategic question which should be asked when designing the IoT system. It is not always about cost and energy efficiency, but also about security and privacy.

%There is no unified framework to design a secure wireless systems based on EH techniques. When the devices are powered by EH sources, they are faced with an additional threat from the attacker who can change the environment. For example the RF source could be blocked and the devices cannot send their data anymore. In fact, the more the devices are geared with the environment, the more they are  vulnerable to the threats. 

We conclude here that researchers need to investigate further to discover the new security threats and the root causes and new IoT features enabled by energy harvesting behind them. We need to design more generic and practical protective measures by taking into account the extra vulnerability added by energy harvesting mechanism. Relying only on energy harvesting techniques is a serious risk which should be carefully evaluated before investing in it. If there is no backup system, EH-enabled systems are not suitable for any application that deals with critical information and require high reliability, availability, and low end-to-end latency.

\begin{figure}[t]
  \centering
  \includegraphics[width=1\columnwidth]{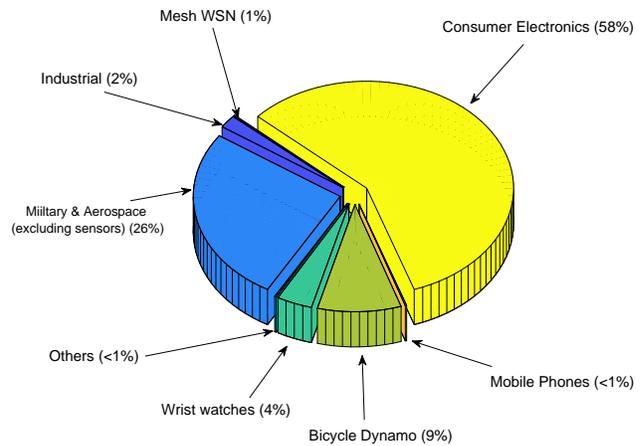}
  \caption{Energy harvesting market 2011 \cite{EHmarket2011}.}
  \label{fig:EHmarket1}
\end{figure}
%\Figure[t](topskip=0pt, botskip=0pt, midskip=0pt)[width=0.98\columnwidth]{EhMarket2017.eps}
%{Energy harvesting market 2017 \cite{EHmarket2011}. \vspace{-6ex}\label{fig:EHmarket2}}
 \begin{figure}[t]
  \centering
  \includegraphics[width=1\columnwidth]{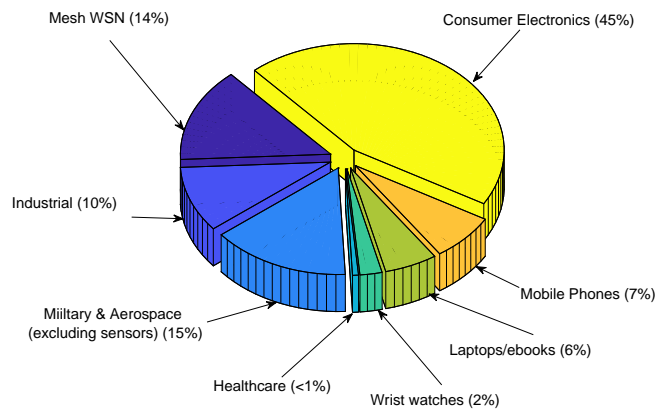}
  \caption{Energy harvesting market 2017 \cite{EHmarket2011}.}
  \label{fig:EHmarket2}
\end{figure}

\section{Energy Harvesting Market Perspective}
%\subsection{Global Energy Harvesting Market}
The global energy harvesting market reached  \$880 million in 2014 and \$1.1 billion in 2015 and will continue to grow to \$4.4 billion by 2021 \cite{bwmarket}. Fig. \ref{fig:EHmarket1} and Fig. \ref{fig:EHmarket2} show the energy harvesting market in 2011 and 2017, respectively. Consumer electronics cover the maximum share of the global EH market.

As shown in Fig. \ref{fig:EHmarket1} and Fig. \ref{fig:EHmarket2}, Industrial and WSNs experience the fastest growth amongst energy harvesting sectors \cite{fminsight}. This is due to the vast deployment of miniaturized devices for industrial automation and monitoring, structural health monitoring, environmental monitoring, and home automation. In other words, the wide adoption of energy harvesting techniques in IoT applications has increased the share of EH market for IoT related sectors, such as industrial and WSNs. It is also important to note that the share of healthcare in EH market is expected to grow very rapidly due to the popularity of IoT applications in healthcare.

%\subsection{Piezoelectric Energy Harvesting  Market}
The market for the piezoelectric energy harvesting is expected to grow significantly which is due to the highest reliability, efficiency and power output by size and cost offered by the piezoelectric energy devices harvesters against the other energy harvesting technologies. The market for industrial applications of piezoelectric energy harvesting is growing significantly due to its wide applications in oil and gas manufacturing, and more generally in industrial environments, where piezoelectric energy harvesting offers a cost-effective alternative to expensive wired infrastructure. Asia Pacific and North American countries are expected to show higher growth in the piezoelectric energy harvesting market over the forecast period \cite{transparency}.

The piezoelectric market can be segmented as industrial switches, consumer electronics, aerospace, healthcare, electronic locks, lighters and other electrical, military, pavements, roads, and railroads, push-button industrial sensors, remote controls, toys and gadgets, and vehicle sensors. Due to potential of piezoelectric nanogenerators for high-tech applications, fabrication of hybrid piezoelectric materials in fiber and powder form open new windows to overcome challenges associate with  applying non-flexible materials in this field \cite{K7}.
%\begin{figure}[t]
%  \centering
%  \includegraphics[width=1\columnwidth]{Piezomarket.jpg}
%  \caption{Global energy harvesting market \cite{EHmarket2011}.}
%  \label{fig:peizomarket}
%\end{figure}
%\Figure[t](topskip=0pt, botskip=0pt, midskip=0pt)[width=0.98\columnwidth]{Piezomarket.jpg}
%{Global energy harvesting market. \vspace{-6ex} \cite{EHmarket2011}.
%  \label{fig:peizomarket}}

One of the main potential use cases of piezoelectric materials is in structural health monitoring. Tiny devices are installed in the structures, such as buildings, bridges, airplanes, and they are expected to continuously work for several years. Powering these devices with batteries is almost impossible as battery replacement is sometime impossible as the devices are out of reach. Battery replacement is not cost-effective and also battery size may be an issue in these applications. Piezoelectric materials provides an unlimited source of energy for these sensors, as structures are rich environments of vibrations. Piezoelectric actuators can be designed in different shapes and sizes; therefore has the potential to lift IoT technologies in structural health monitoring.
%\begin{figure*}
%  \centering
%  \includegraphics[width=1.4\columnwidth]{EnergySourcesIoT.eps}
%  \caption{Energy sources for IoT applications.}
%  \label{fig:taxonomy}
%\end{figure*}

Healthcare industry is also considered as one of the main potential targets for piezoelectric materials. Implanted devices can be powered by piezoelectric materials instead of batteries, to increase the lifetime of these devices and make them comfortable for the patients. Wearable sensors, such as smart watches and fitness trackers, can be also powered by piezoelectric materials to convert human body motion into electricity to perform measurements and wireless communication with personal devices.

%\begin{figure}[t]
%  \centering
%  \includegraphics[width=1\columnwidth]{Figx3.png}
%  \caption{Applications of piezoelectric nanogenerators under different loading conditions \cite{crossley2015energy}.}
%  \label{fig:X3}
%\end{figure}

Some of the key players in the piezoelectric energy harvesting market include Advanced Cerametrics, Boeing, Honeywell, ITT, Microstrain, Inc., Smart Material Corp., and Tokyo Institute of Technology \cite{transparency}. As reported by IDTechEx, about \$145 million in 2018 will be spent on piezoelectric energy harvesting and by 2022 it will create a market of \$667 million \cite{idtech}.

%\label{Sec-Intro}

\section{Conclusions}
This paper reviews energy harvesting techniques for Internet of Things (IoT) services and applications. Over 50 billion multi-role devices, capable of sensing and actuating, will be installed by 2025, which shows a tremendous growth in the number of devices and creates new challenges and opportunities. A major burden is powering these devices, as using the main power and batteries is mostly restricted due to the small sizes of many devices and the fact that these devices are installed in hard-to-reach areas, where regular battery maintenance is impractical and very expensive. A viable solution is to use energy harvesting techniques to harvest energy from environment and provide enough energy to the devices to perform their operations. This will significantly increase
the device life time and eliminate the need for the battery as an energy source. Different energy harvesting techniques were presented in this survey and pros and cons of each technique were discussed. As efficient energy harvesting technique, we focused on piezoelectric energy harvesting and radio frequency energy harvesting due. We briefly introduced the main concepts  and design challenges for these technologies. As short-range wireless technologies are operating at mW power range, the development of battery-less IoT devices may be feasible. However, due to the large transmit power of most LPWAN technologies, which are expected to play key roles to provide massive IoT services, batteries will remain an essential parts of IoT devices operating over LPWANs. Energy harvesting techniques will then play key roles in increasing the device life-time by providing  a sustainable way to recharge the batteries.

\bibliographystyle{IEEEtran}
\footnotesize
\bibliography{IEEEabrv,References}
%
%\vskip-8ex
\begin{IEEEbiography}[{\includegraphics[width=1in,clip]{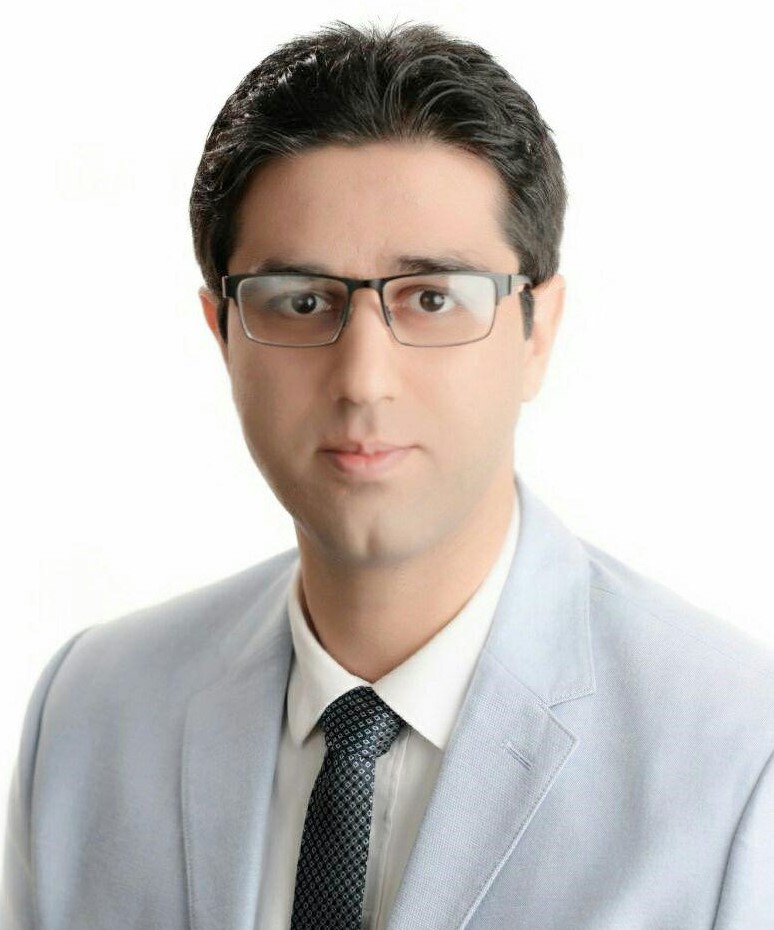}}]{Mahyar Shirvanimoghaddam} (S'12-M'15) received the B.Sc. degree (Hons.) from the University of Tehran, Iran, in 2008, the M.Sc. degree (Hons.) from Sharif University of Technology, Iran, in 2010, and the Ph.D. degree from The University of Sydney, Australia, in 2015, all in electrical engineering. He held a post-doctoral research position with the School of Electrical Engineering and Computer Science, University of Newcastle, Australia. Since 2016, he has been with the School of Electrical and Information Engineering, The University of Sydney, as a Scholarly Teaching Fellow in telecommunications. His general research interests include channel coding techniques, cooperative communications, compressed sensing, machine-to-machine communications, and wireless sensor networks. He was selected as an Exemplary Reviewer of the IEEE Communication Letters and the IEEE Transaction on Communications.
\end{IEEEbiography}
%\vskip-9ex
\begin{IEEEbiography}[{\includegraphics[width=1in,clip]{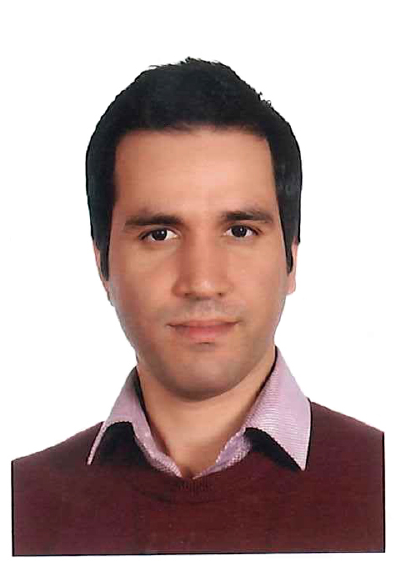}}]{Kamyar Shirvanimoghaddam} is a PhD candidate in Institute for Frontier Materials, Deakin University, Australia.  He worked as a materials scientist in different companies and universities since 2004 when he graduated from University of Tehran,Iran. As a doctoral candidate, he is focusing on develop a new periodical patterned CNT for high tech applications such as sensors and nanogenerators .His expertise includes Carbonaceous structure, composite materials, and piezoelectric polymers.
\end{IEEEbiography}
%\vskip-10ex
\begin{IEEEbiography}[{\includegraphics[width=1in,clip]{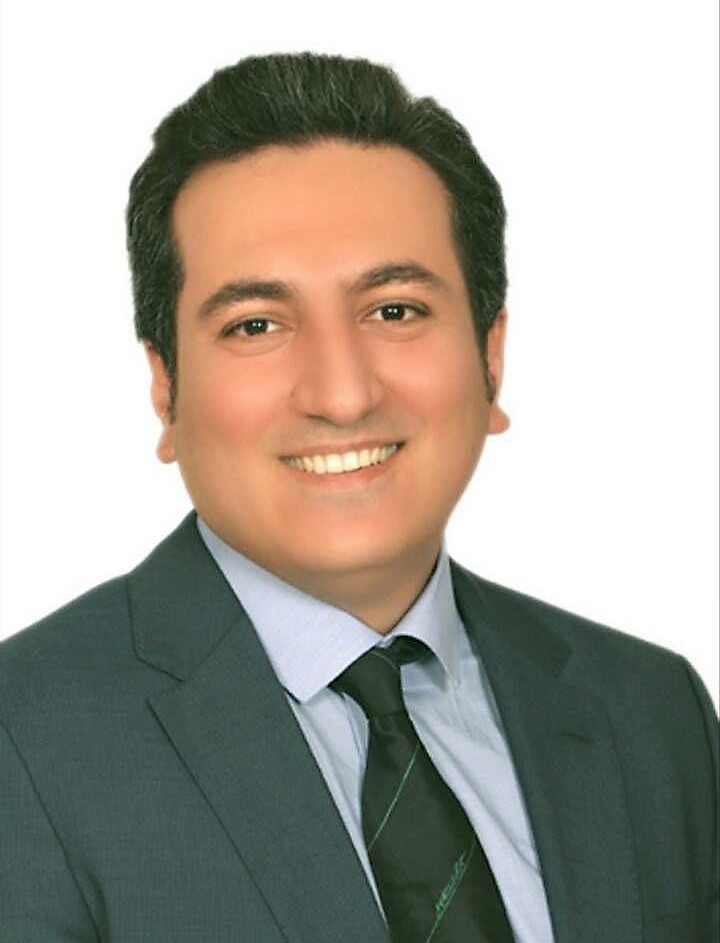}}]{Mohammad Mahdi Abolhasani}  received his PhD degree in polymer engineering from Amirkabir University of Technology, Tehran,Iran. His PhD research work contributed to the demonstration of effects of nanofillers and dynamic vulcanization on crystalline polymorphism of PVDF. Then, he joined University of Kashan, Iran, as an assistant professor. His work at Kashan was focused on investigation of piezoelectric performance of PVDF films and fibers. In 2014, he did a Postdoc in Deakin University, Australia. In 2016, he received a fellowship from the Alexander von Humboldt Foundation for a project on high-K ferroelectric polymers for energy storage application. He is now a Postdoc researcher in Max-Planck institute for polymer research, Germany. 
\end{IEEEbiography}
\vskip-6ex
\begin{IEEEbiography}[{\includegraphics[width=1in,clip]{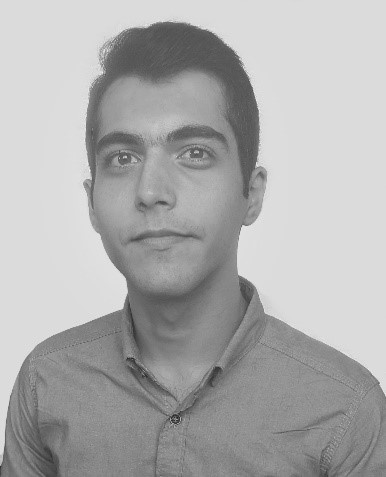}}]{Majid Farhangi} received the B.Sc. degree in Electrical Engineering from the IAUCTB, Tehran, Iran, in 2015. Currently, he is pursuing the Master of Electrical Engineering degree at the University of Sydney, Australia. His current research interests mainly focus on high-efficiency DC-DC converters, wireless power transmission and high frequency induction heating systems.
\end{IEEEbiography}
\vskip-6ex
\begin{IEEEbiography}[{\includegraphics[width=1in,clip]{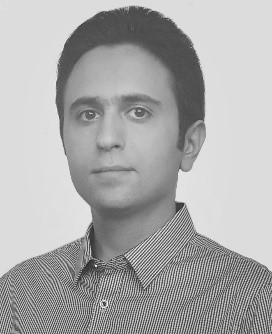}}]{Vahid Zahiri Barsari}  received the B.S. degree in electrical engineering from IAUCTB, Tehran, Iran, in 2015. Currently, he is pursuing the Master of Electrical Engineering degree at the University of Sydney, Australia. His current research interests mainly focus on power electronics, power inverter and designing inductive power transfer systems for electric vehicles.
\end{IEEEbiography}
\vskip-6ex
\begin{IEEEbiography}[{\includegraphics[width=1in,clip]{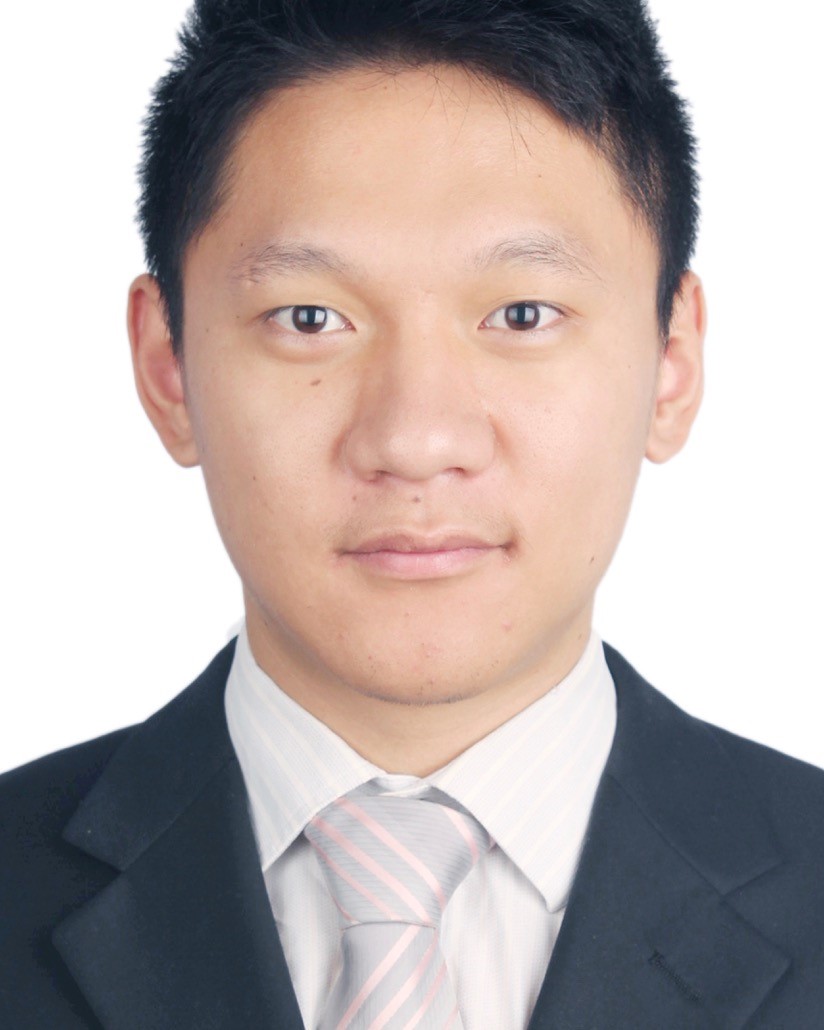}}]{Hangyue Liu} (S'16) received the B.E. degree from SouthWest Jiaotong University, Chengdu, China, in 2014 and the M.E. degree in electrical engineering from the University of Sydney, Sydney, Australia, in 2017. He is currently a research assistant with the School of Electrical and Information Engineering, University of Sydney, Sydney, Australia. His current research interests include demand response, energy management in residential buildings, and applications of IoT in smart grid.
\end{IEEEbiography}
\vskip-6ex
\begin{IEEEbiography}[{\includegraphics[width=1in,height=1.25in,clip]{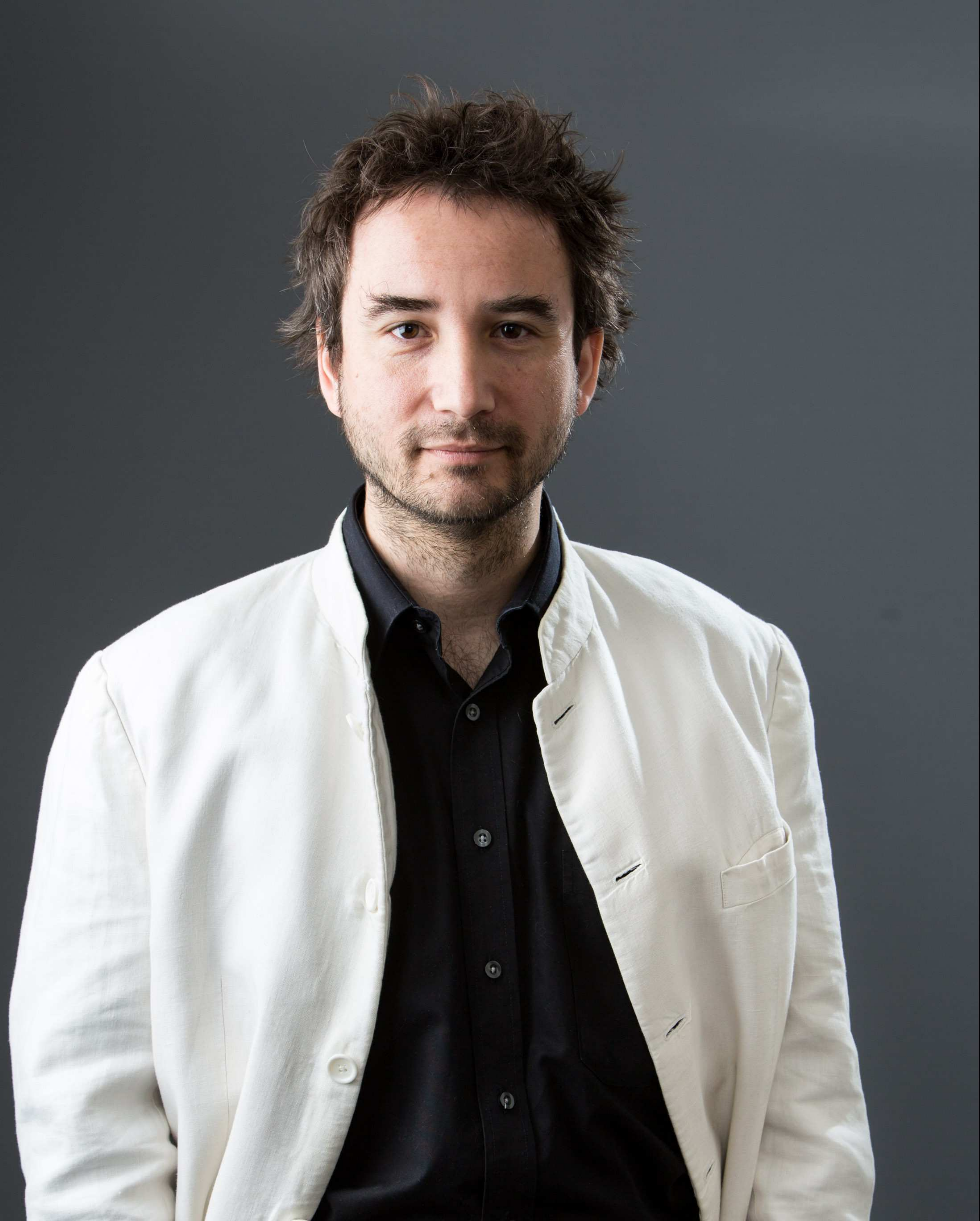}}]{Mischa Dohler} (S'99-M'03-SM'07-F'14) is currently a Full Professor in Wireless Communications with King’s College London, the Head of the Centre for Telecommunications Research, and a co-founder and a member of the Board of Directors of the smart city pioneer Worldsensing. He was a Distinguished Lecturer of the IEEE. He was the Editor-in-Chief of the Transactions on Emerging Telecommunications Technologies and the Transactions on the Internet of Things.

He is a frequent keynote, panel and tutorial speaker, and has received numerous awards. He has pioneered several research fields, contributed to numerous wireless broadband, IoT/M2M and cyber security standards, holds a dozen patents, organized and chaired numerous conferences, has more than 200 publications, and authored several books.

He acts as policy, technology and entrepreneurship adviser, examples being Richard Branson’s Carbon War Room, House of Parliament UK, UK Ministry BIS, EPSRC ICT Strategy Advisory Team, European Commission, Tech London Advocate, ISO Smart City working group, and various start-ups. He is also an entrepreneur, angel investor, passionate pianist and fluent in 6 languages. He has talked at TEDx. He had coverage by national and international TV \& radio; and his contributions have featured on BBC News and the Wall Street Journal.
\end{IEEEbiography}
\vskip-6ex
\begin{IEEEbiography}[{\includegraphics[width=1in,height=1.25in,clip]{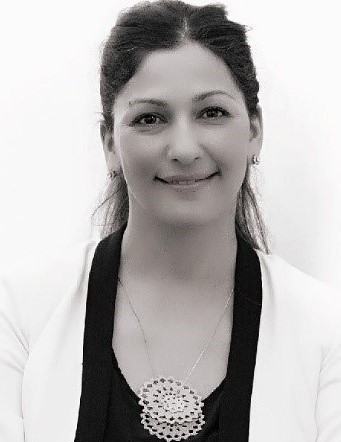}}]{Minoo Naebe} received the the B.Sc. (Hons.) and M.S  degrees from the Isfahan University of Technology (IUT), Isfahan, Iran, and the Ph.D. degree  from Deakin University, Australia (2008). She is a Senior Research Fellow and the Theme Leader for High Performance Materials with Carbon Nexus, Institute for Frontier Materials, Deakin University. Her research interests include advanced polymer (nano) composites and fibres and applications in various industries  aerospace, automotive, oil and gas.
\end{IEEEbiography}
\end{document}